\documentclass{pasj00}
\draft  

\begin{document}
\SetRunningHead{L. Zaninetti }{Shape of Superbubbles in the Galactic 
Plane}
\Received{2003/12/4}
\Accepted{2004/10/27}

\title{On the Shape of Superbubbles Evolving in the \\ Galactic Plane}

\author{Lorenzo Zaninetti  }

\affil {Dipartimento di Fisica Generale, Via Pietro Giuria 1,\\
           10125 Torino, Italy}

\email   {zaninetti@ph.unito.it}

\KeyWords{ISM: bubbles -- ISM: clouds -- Galaxy: disk -- galaxies: starburst -- methods: numerical }

\maketitle

\begin{abstract}

The galactic  supershells  are cavities  in  the
interstellar  medium.
These  shells  can be explained  by introducing the concept
of superbubbles, the theoretical result 
of  multiple supernova.   
The superbubbles   can  be analytically
described  if the ambient medium has a constant density
both in the so-called bursting phase
and  in the subsequent adiabatic expansion.
In order to solve   the expansion of  superbubbles 
in the ISM, which   is  a
non-homogeneous  medium, a  numerical technique is used that 
divides the sphere  into many sectors.
By varying the time of the bursting phenomenon and  the time
over which  the phenomenon is followed,   elliptical and
hour-glass shapes or vertical
walls can be obtained.
Application of the developed theory/code
to the super-shell  associated with GW~46.4+5
and  with GSH~238 allows us to say that  the suggested
physical parameters   are consistent with our theory.
The map of the expanding superbubble's velocity can
 be   tentatively traced  by
generating  random points on the expanding surface.
The structure of the galactic plane as a result
of  the evolution of many super-bubbles
was simulated by adopting   the 
percolation theory in order
to   generate new OB associations.
\end{abstract}

\section{Introduction}

The term  supershell  was observationally defined by \citet{heiles1979}
where eleven H\,I objects were examined.
The  supershells   have been observed  as   expanding  shells,
or holes,  in the H\,I-column density   distribution  of  our galaxy,
in the   Magellanic Clouds,
in   the dwarf irregular
galaxy Ho-II~\citep{puche},
and in the nearby  dwarf galaxy IC 2574~\citep{walter}.
The dimensions of these objects  span from 100~\mbox{pc}
to 1700~\mbox{pc}
and often present  elliptical shapes or elongated features, which
  are difficult to explain based on   an expansion in a uniform
medium.
These structures  are commonly  explained through introducing
theoretical objects named  bubble or  superbubble;
these  are created by  mechanical energy input from stars 
(see  for example~\cite{pikelner} ; \cite{weaver}).
Thus, the origin of a supershell is not necessarily a 
   superbubble, and other
   possible origins include collisions of high-velocity clouds (see
   for example~\cite{tenorio} ; \cite{santillan}~).
   The  worm is another 
    observed feature that may, or may not, be associated with a wall
    of a superbubble.
    Galactic worms were 
    first identified as irregular, vertical columns of
    atomic gas stretching from the galactic plane;
    now, similar structures are found
    in radio continuum and infrared maps (see for example~\citet{koo}). 

The models 
that  explain  supershells as being due to 
the combined  explosions  of supernova in a cluster  of
massive stars are now briefly reviewed.
Semi-analytical and hydrodynamical
calculations are generally adopted.
In  semi-analytical calculations
the thin-shell approximation
can be the key to obtaining
the expansion of
the superbubble; see,  for example,
\cite{mccray,mccrayapj87,mac0,igu,basu}.

   The thin-shell approximation has already been  used in a variety of
   different problems, and with both analytical and numerical approaches
   (aside from the review by \cite{bisnovatyi}, see \cite{begelman}; 
    \cite{moreno}). Thus 
   the validity and limitations of the method are well known. For instance,
   modelling is fast and simple because only shell dynamics are included
   in an approximate way, while fluid variables are not included
   at all. 
  The price
  that one has to pay is a lack of knowledge of  the density and
  velocity profiles and, obviously,  the onset and
  development
  of turbulence, instabilities or mixing  cannot be followed. 
  These facts limit the applications
  of this method to derive only general shapes, approximate expansion
  rates, and gross features of the surface mass-density distributions 
of the
  shells.
  A detailed comparison with the results of so-called thin-layer 
  approximation, as described for example in
  \citet{silich4}  is now being carried out.
  \begin{itemize}
   \item 
   The two main assumptions, which are (i)  all swept-up gas 
   accumulates infinitely in a thin shell just after
   the shock front and (ii) the gas pressure is uniform
   inside the cavity,  are in common with the 
   two models.
   \item
    The numerical equations of  motion  are different. 
    For examples~(\ref{eq_momentum}) and  (\ref{eqn:dpdt})
    are for during the bursting phase,  and Silich's set  
    from (5) to (9) is for  the three Cartesian coordinates.
    \item In  our  model  the  shear due to  
      galactic rotation  is neglected in a first approximation.
      See  subsubsection~\ref{galrot} for a  late addition 
      on  shear. 
    \item Our treatment includes the concept of efficiency 
         of the simulation, equation (\ref{eq:reliability}),  
         which  is neglected by Silich.    
    \end {itemize}
Another  source of comparison can be made with the 3D models
of~\citet{silich}, in which     a number of processes not considered 
here are now included:
\begin {itemize}
\item  Two mechanisms of mass injection into the cavity 
       are adopted: thermal evaporation of both 
       the cold expanding shell, and the  clouds engulfed 
       by the shell,  and dynamic disruption 
       of clouds penetrating into the bubble interior.
\item  A power-law density and temperature distribution 
       inside the superbubble during the entire evolution.
\item  A  cooling function that takes into  account 
       the metallicity gradients in the Galaxy.
\end {itemize}

However  the astrophysical results  of~\citet{silich}
are basically similar to  ours, and  the maximum elongation
of the superbubbles along the z-direction can reach 2~kpc 
in  both models.
Similar  results   are also obtained concerning
the morphologies: in both models the hourglass-shape is 
obtained when   equality between  the 
 bursting time    and  the elapsed time
is assumed. The  vertical walls  considered here are
 the belt  of subsection 4.2  of ~\citet{silich}.

The hydrodynamical approximation was used by \citet {mac}.
As for the effect of magnetic fields,
a  semi-analytical method was  introduced
by~\citep{ferriere} and
a  magneto-hydrodynamic 
code has been adopted  by various authors
\citep{tomisaka2,tomisaka3,kamaya}.

The plane of the numerical analysis is now outlined.

A numerical code that solves the momentum equation coupled
with  the variation of pressure in the presence
of the injection of
mechanical luminosity and adiabatic losses was  developed
in section~\ref{code};  testing with  numerical
hydro-dynamic 
 calculation was developed in subsubsection~\ref{sec:hydro}.

The  analytical ``expansion law''  of the superbubbles
can   be
set up both in the bursting phase ( see section~\ref{approximation})
and  in the adiabatic phase after the SN burst stops,
(see subsection~\ref{approximation2}).

The various types of obtained shape 
(elliptical,egg,V, vertical  and hourglass) 
 are described in subsection~\ref{shapes}.

A numerical law  that describes the raising of the vertical
walls is deduced in subsubsection~\ref{verticalwalls}.

The map of the expanding superbubble's velocity can be
tentatively traced by generating
random points on the expansion
surface ( see  subsection~\ref{ref_velocity}).

The developed theory was
then applied  to the supershells associated with GW~46.4+5~
and GSH-238 (see subsection~\ref{sec464} and 
 subsection~\ref{sec238},
 respectively).

The H\,I structure in  the
galactic plane
can be tentatively simulated
by using the percolation framework to trace the
spiral-arm  structure
( see section~\ref{percolation}
and  appendix~\ref{spiral}).

By adopting the framework of radiation coming from
the layer comprised between two shells,  we can
deduce the  center darkening law (see appendix~\ref{sec_ring}).

\section{The Approximation Used}
\label{code}
In our case, the starting equation for the evolution
of the superbubble~\citep{mccray,mccrayapj87}
 is  momentum conservation
 applied to a pyramidal section,   characterised
by a solid angle, $\Delta \Omega_j$:
\begin {equation}
\frac{d}{dt}\left(\Delta M_j \dot{R}_j\right)=p R_j^2
\Delta \Omega_j
,
\label{eq_momentum}
\end {equation}
where the pressure of the surrounding medium is
assumed to be negligible and the mass
is confined into a thin shell with mass $\Delta M_j$.
The subscript $j$  was added  here
in order to note
that this is not a spherically symmetric system.
Due to the fact that  $p$ is uniform in the cavity of
the superbubble, a summation to obtain the
total volume, such as
\begin{equation}
V=\sum  \Delta \Omega_j R_j^3/3
\label{eq_volume}
,
\end  {equation}
is necessary  to determine its value.
The mass conservation equation for a thin
shell is
\begin {equation}
\Delta M_j = \frac{1}{3}R_j^3 \bar{\rho}_j\Delta\Omega_j
 .
\end {equation}
The  pressure is enclosed in the energy conservation equation,
\begin{equation}
\frac{1}{\gamma-1}\frac{d(pV)}{dt}=L-p\frac {dV}{dt},
\label{eqn:MMenergy}
\end{equation}
where $L=E_0R_{\rm SN}/4\pi$ denotes
the mechanical luminosity
injected into  a
unit
solid angle and $\gamma~=5/3$.
Equation~(\ref{eqn:MMenergy}) can be expanded,
obtaining
\begin {equation}
\frac {dp}{dt} = \frac { L (\gamma -1 )} {V}
            - \gamma  \frac {p}{V} \frac {dV}{dt}
\label{eqn:dpdt}
 .
\end {equation}
Formulae~(\ref{eq_momentum}) and  (\ref{eqn:dpdt})
will be our basic equations to  numerically integrate 
in the bursting phase.
An approximation concerning the pressure can  be obtained by
ignoring the second  term of the rhs in~(\ref{eqn:MMenergy});
this leads to
\begin {equation}
p=\frac{2E_0R_{\mathrm{SN}}t}{3V}
,
\label {eq:pressure}
\end {equation}
where  t is the considered time, $R_{\mathrm{SN}}$
the rate of supernova explosions, $E_0$ the energy of each
supernova, and V   is computed as in  equation~(\ref{eq_volume}).
On   continuing  to   consider   the case of constant density,
the volume becomes:
\begin {equation}
V=\frac{4\pi}{3}R^3
 .
\end {equation}
Equations (\ref{eq_momentum}) and (\ref{eq:pressure})
 lead to
\begin {equation}
\frac{d}{dt}\left(\bar{\rho}R^3\dot{R}\right)=
\frac{3E_0R_{\mathrm{SN}}}{2\pi}
\frac{t}{R}
 .
\label {eq:firstder}
\end {equation}
$R^3 \dot{R}=AR^\alpha$
is imposed to integrate this equation.
After adopting the initial condition of  $R=0$ at $t=0$
and assuming
$\bar{\rho}$ is constant irrespective of $R$ or $t$,
the equation
on the expansion speed is obtained

\begin {equation}
\dot{R}=\frac{\alpha+1}{\alpha}
 \frac{3E_0R_{\mathrm{SN}}}{4\pi \bar{\rho}R^4}t^2
 .
\end {equation}
\label{approximation}

By integrating  the previous equation,
 ``the expansion law'' is obtained,
\begin {equation}
R=\left[\frac{5(\alpha+1)}{4\pi \alpha}\right]^{1/5}
  \left(\frac{E_0R_{\mathrm{SN}}}{\bar{\rho}}\right)^{1/5} t^{3/5}
 ,
\label{eq:firstradius}
\end {equation}
which is identical to  equation (10.34) of
~\citet{mccray},
 since $\alpha=7/3$.

\subsection{After SN Bursts Stop}

\label{approximation2}
It is clear that  an upper limit
should be inserted into
the basic equation~(\ref{eq:pressure}); this  is the time
after which the bursting phenomena stops, ${t^{\mathrm{burst}}}$.

Since there is the $pdV$ term in the the first theorem
of thermodynamics ($dQ=0=dU+pdV$),
the total thermal energy decreases with time.
The pressure of  the internal gas decreases according to
the adiabatic law,
\begin {equation}
p=p_1({V_1 \over V})^{5/3}
\label {eq:adiabatic1}
,
\end {equation}
where
\begin {equation}
p_1 =  \frac {2}{3}  \frac {E_0R_{\mathrm{SN}} t^{\mathrm{burst}}}
    {  V_1}
,
\label {eq:adiabatic3}
\end {equation}
and the equation for
the conservation of  momentum
becomes
\begin {equation}
\frac{d}{dt}\left(\bar{\rho}_j~R_j^3\dot{R_j}\right)=
3 \frac{p_1 V_1^{5/3} }{ V^{5/3}   }
R_j^2
 .
\label {eq:adianum}
\end {equation}
It is important to remember that this phase occurs after the SN burst stops.
At that time ($t=t^{\mathrm{burst}}$), the volume  of the bubble
is computed  by using
$R_j(t^{\mathrm{burst}})$,
 and the expansion speed is equal
to $\dot{R}_j(t^{\mathrm{burst}})$,
both of which are obtained from
the numerical solution of
equation (\ref{eq:firstder}).
On assuming  (also here)   that
$\bar{\rho}$ is constant irrespective of $R$ or $t$,
equation~(\ref{eq:adiabatic3}) is now inserted into
equation~(\ref {eq_momentum}) and
  the expansion law  is obtained for the after-burst phase,
\begin {equation}
R=\left(\frac{147} {4\pi }\right)^{1/7}
  \left(\frac{E_0R_{\mathrm{SN}} t^{\mathrm{burst}} R_1^2 }
    {\bar{\rho}}\right)^{1/7} t ^{2/7}
 ,
\label{eq:secondstradius}
\end {equation}
which is identical to  equation (10.33) of
~\citep{mccray}.

\subsection  {Numerical Integration}

\label{sectheta}
The differential  equations need to be solved by sectors,
with each sector being treated  as independent from the others,
except for the coupling in computing 
 the  volume $[$ see equation~(\ref{eq_volume})$]$ of the bubble.
From a practical point of view,  the
range   of the polar angle $\theta$ ($180^{\circ}$)
will be divided
into  $n_{\theta}$ steps, and the range of the azimuthal angle
$\phi$ ($360^{\circ}$) into  $n_{\phi}$   steps.
This will  yield  ($n_{\theta}$ +1) ($n_{\phi}$ +1) directions of
motion that can  also  be identified with  the number of vertices
of the polyhedron representing the volume occupied by the
expansion;
this   polyhedron  varies  from  a sphere
to various morphological  shapes  based on  the
swept up  material
in each direction.

In  3D plots showing the  expansion surface  of the explosion,
(see for example figure~\ref{worm}),
 the number  of vertices is    $n_{\mathrm {v}}$ =
\label{sec_nv}
($n_{\theta}$+1)$\cdot$($n_{\phi}$ +1)
and  the number
of  faces is 
$n_{\theta}\cdot\;n_{\phi}$,
typically we have     $n_{\theta}$=50
and $n_{\phi}$=50 (in this case the subscript  varies
between 1 and 2601).
However, all calculated  models  are axisymmetric,
and the essential number of points to draw such
figures is only $n_\theta+1$=51.
 $R_{\mathrm{up}}$,
$R_{\mathrm{eq}}$ and  $R_{\mathrm{down}}$  are now introduced, which
represent  the distances  from the  position of  the
OB associations  (denoted by $z_{\mathrm{OB}}$)
to the top, to the left and  to
the bottom of the bubble;
see,  for example,
figure~\ref{confronto} where $z_{\mathrm{OB}}$ =100~pc.
\label{sec:radii}

\subsection {The Numerical Equations}

At each  time step ,$\Delta t$,   the volume {\it V}  of the
expanding bubble is computed $[$see equation~(\ref{eq_volume})$]$.
In  other words,    the volume {\it  V }
swept up   from the explosions  is no  longer a sphere,
but  becomes an egg or an hourglass.
The pressure in the first phase,
see equation~(\ref{eqn:dpdt}),  is computed through the
following finite-difference approximation:
\begin{equation}
p^k  =
p^{k-1} + \left [  L \frac {\gamma -1} {V^k}
- \gamma \frac {p^{k-1}}{V^k}  \frac{V^k -V^{k-1}} {\Delta t}
\right ]  \Delta t
,
\end{equation}
where k is the number of steps considered.

Equation~(\ref{eq_momentum})
now leads  to
\begin {equation}
\frac{d}{dt}\left(\bar{\rho_j} R_j^3\dot{R_j}\right)=
3 pR^2_j
 ,
\label {eq:firsnum}
\end {equation}
which  may  rewritten as
\begin {equation}
3 R_j^2 \dot{R_j}^2 + R_j^3 \ddot {R_j} =
3p \frac{R^2_j}{\bar{\rho_j}}
- \frac{\dot{ \bar{\rho_j}} }{\bar{\rho_j}} R_j^3\dot{R_j}
 .
\label {eq:secondnum}
\end {equation}
The first term represents the ram pressure of the stratified ISM
on the expanding surface and the second represents the inertia
of the bubble. The average density $\bar{\rho_j}$ is
numerically computed according to the algorithm outlined
in subsection~\ref {sec_rhoaverage}; the time derivative of
$\bar{\rho_j}$ at each time step $\Delta t$
 is computed according
to the finite difference-approximation,
\begin{equation}
\dot{ \bar{\rho_j}} =
\frac { \bar{\rho_j}^k -\bar{\rho_j}^{k-1}}
{\Delta t}
,
\label{deriv_rho}
\end{equation}
where {\it k } is the number of steps considered.

Equation~(\ref{eq:secondnum})  can be re-expressed
in two  differential
equations (along each direction j)
of the first order suitable to be integrated:
\begin {equation}
 \frac {dy_{1,j}} {dt}   = y_{2,j}
,
\end  {equation}
\begin{equation}
\frac  {dy_{2,j}}{dt}    =
3p \frac{ 1  }{ \bar{\rho_j} y_{1,j}}
-3 \frac { y_{2,j}^2 }{ y_{1,j}}
- \frac{\dot{ \bar{\rho_j}} }{\bar{\rho_j}} y_{2,j}
 .
\label {eq:twonum}
\end {equation}

In the new  regime ($t \geq t^{\mathrm{burst}}$)
equation~(\ref{eq:adianum}) now becomes
\begin {equation}
3 R_j^2 \dot{R_j}^2 + R_j^3 \ddot {R_j} =
3 \frac{p_1 V_1^{5/3} }{ V^{5/3}   \bar{\rho_j}}
R_j^2
- \frac{\dot{ \bar{\rho_j}} }{\bar{\rho_j}} R_j^3\dot{R_j}
 .
\label {eq:numericaladia}
\end {equation}
Equation~(\ref{eq:numericaladia})  can be re-expressed
as  two  differential
equations (along each direction {\it j})
of the first order suitable to be integrated:
\begin {equation}
 \frac {dy_{1,j}}{dt}   = y_{2,j}
,
\end  {equation}
\begin{equation}
\frac  {dy_{2,j}}{dt}   =
\frac{3 p_1 V_1^{5/3}} { V^{5/3} \bar{\rho_j} y_{1,j}}
-3 \frac { y_{2,j}^2 }{ y_{1,j}}
- \frac{\dot{ \bar{\rho_j}} }{\bar{\rho_j}} y_{2,j}
 .
\label {eq:twonumadia}
\end {equation}

The  integrating scheme  used  is a  Runge-Kutta  method,
and   in particular the  subroutine RK4
(\cite{press}); the  time  derivative of the  density
along a certain direction j, $\dot{ \bar{\rho_j}}$,
is computed at each time step
according to formula~(\ref{deriv_rho}).

The integration time, $t_{\mathrm{age}}$,  and the time steps
are always  indicated in the  captions
connected with the various diagrams.

From a numerical point of view,  the pressure
across the bursting time  
is continuous. 
This  is because the first value  of the pressure
after the bursting time  is that  of the last step
in the bursting phase,  modified according to the  
adiabatic law   modelled  by  formula~(\ref{eq:adiabatic1}).

The upper limit  chosen to integrate the  differential 
equations is  $t_{\mathrm {age}}=2.5\cdot10^7~{\mathrm{yr}}$;
this is the approximate age of GSH~238, the 
super-shell simulated  in subsection~\ref{sec238}.

\subsection{The Density Profile}

\begin{figure}
  \begin{center}
\FigureFile(120mm,120mm){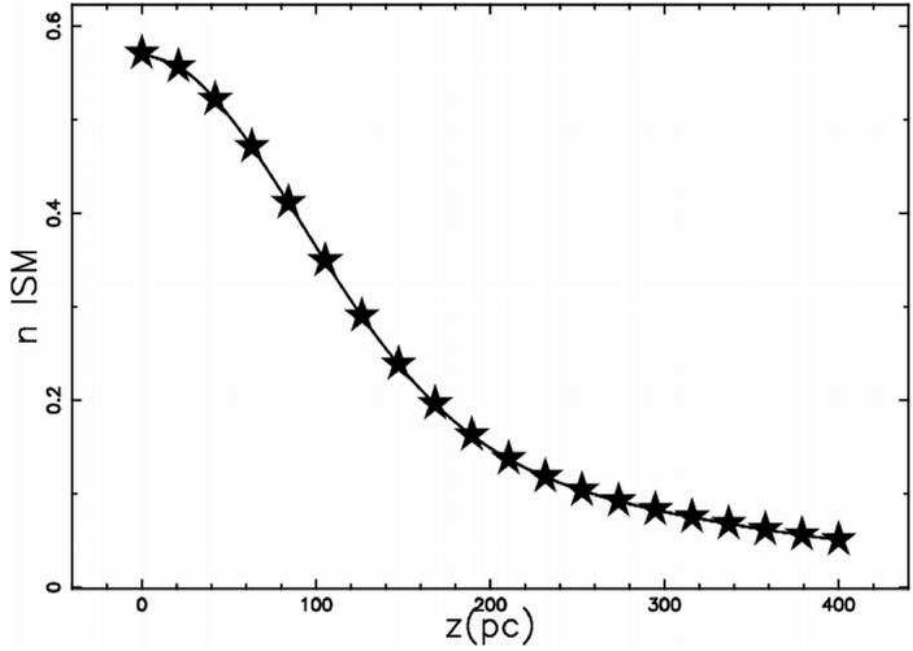}
  \end{center}
\caption{
Average structure of the gaseous disk in the {\it z}  -direction;
{\it z}  is allowed to vary between 0 \mbox{pc} and  400 \mbox{pc}.
}
\label{figism}%
    \end{figure}

The vertical density distribution
of galactic H\,I  is well-known; specifically, it has the following
three component  behaviour as a function of
{\it z} , the  distance  from  the galactic plane in pc:
\begin{equation}
n(z)  =
n_1 e^{- z^2 /{H_1}^2}+
n_2 e^{- z^2 /{H_2}^2}+
n_3 e^{-  | z |  /{H_3}}
\,.
\label{equation:ism}
\end{equation}
\label{Sec_ISM}
We took~\citep{bisnovatyi,dickey,lockman} 
 $n_1$=0.395 ${\mathrm{particles~}}{\mathrm{cm}^{-3}}$, $H_1$=127
        \mbox{pc},
        $n_2$=0.107 $\mathrm{particles~}{\mathrm{cm}^{-3}}$, $H_2$=318
        \mbox{pc},
        $n_3$=0.064 $\mathrm{particles~}{\mathrm{cm}^{-3}}$, and  $H_3$=403
        \mbox{pc}.
This  distribution  of  galactic H\,I is valid in the range
0.4 $\leq$  $R$ $\leq$ $R_0$, where  $R_0$ = 8.5 \mbox{kpc}
and $R$  is the
distance  from  the galaxy center.
A plot  showing such a dependence of  the ISM 
density 
from
{\it z}  is shown in figure~\ref{figism}.

\subsection{Computation of the Swept Mass}

The  ISM density is not constant, but  varies  in the
{\it z}-direction according  to  equation~(\ref{equation:ism}).
The swept mass can therefore be computed in a
certain direction {\it j}  using the
following  algorithm:
\begin{enumerate}
\item
The pyramidal sector is divided into layers (for example 1000)
whose radii range from $R_{L-1/2}$ to $R_{L+1/2}$.
\item  In each layer  the volume
$
\Delta V_L=\frac{1}{3}(R_{L+1/2}^3-R_{L-1/2}^3)
$
is computed as well  as the corresponding   mass,
$\Delta M_L=\Delta V_L\rho_j(z=R_L\cos \theta_j+z_{\mathrm{OB}})$,
where $\theta_j$ represents an angle between the $z$-axis and the $j$-th
radial path.

\item The various contributions, $\Delta M_L$,
are added in order to obtain
the total swept  mass in the considered sector.

\item
The average density along a sector {\it j}
 as $\bar{\rho_j}$ is calculated
 using $\Delta M_L$ and $\Delta V_L$ as $\bar{\rho_j}=\sum_L \Delta M_L/
\sum_L\Delta V_L$.

\end{enumerate}

\label {sec_rhoaverage}

\subsection  {Astrophysical Units}

Our basic units are: time ($t_7$), which
is expressed  in $10^7$ \mbox{yr} units;
$E_{51}$, the  energy in  $10^{51}$ \mbox{erg};   $n_0$  the
density expressed  in particles~$\mathrm{cm}^{-3}$~
(density~$\rho_0=n_0$m, where m=1.4$m_{\mathrm {H}}$); and
$N^*$,  which  is the number of SN explosions
in  $5.0 \cdot 10^7$ \mbox{yr}.

By using the previously defined  units,
formula~(\ref {eq:firstradius}) concerning
``the expansion law'' in the bursting phase  becomes
\begin{equation}
R =111.56\;  \mathrm{pc}(\frac{E_{51}t_7^3 N^*}{n_0})^{\frac {1} {5}}
.
\label{eqn:raburst}
\end{equation}

Conversely, equation~(\ref {eq:secondstradius}) concerning
  ``the expansion law'' in the adiabatic phase
is
\begin{equation}
R =171.85\;  \mathrm{pc}( \frac {E_{51}N^*} {n_0} )^{\frac {1}{5}}
                   (t^{\mathrm{burst}}_7)^{\frac {11}{35}}
                   (t_7)^{\frac {2} {7} }
 .
\label{eqn:raadia}
\end{equation}
This is the approximated  radius derived by a spherical model.
A perfect coincidence is not expected
upon  making  a comparison  with
that expected in
a stratified medium obtained with the multidimensional
thin shell
approximation.
Another useful formula is  the luminosity of the bubble
(see~\cite{mccray}),
\begin{equation}
L_{\mathrm{SN}}=E_0R_{\mathrm{SN}}=0.645^{36} \mbox {erg}~  \mbox{s}^{-1}E_{51}N^*
  .
\end{equation}
This formula can be useful in order to derive
the parameter  $N^*$ based on  observations;
 for example,
a luminosity of  $1.6~10^{38} \mbox {erg~s}^{-1}$
corresponds to  $N^*=248$.  
The total deposited energy, $E_{\mbox{tot}}$, is
\begin{equation}
E_{\mbox{tot}} = E_{51}N^* 10^{51} 
\frac {t_{\mathrm {age}}}{5 \cdot 10^7 {\mathrm{yr}}} {\mbox{erg}}
,
\label{etotage}
\end{equation}
when $t^{\mathrm{burst}}$=$t_{\mathrm {age}}$ and
\begin{equation}
E_{\mbox{tot}} = E_{51}N^* 10^{51} 
\frac {t^{\mathrm{burst}}}{5 \cdot 10^7 {\mathrm{yr}}} \mbox{erg}
,
\label{etotburst}
\end{equation}
when $t^{\mathrm{burst}}~<~t_{\mathrm {age}}$.

The spectrum of  the radiation emitted depends on
 the temperature
behind the shock front $[$ see  for example formula 9.14  
in~\citet{mckee}$]$,
\begin{equation} 
T = \frac{3}{16} \frac{\mu} {k} v_{\mbox{s}}^2  ~~~{\mbox{K}}^{\circ}
,
\end{equation}
where $\mu$ is the mean mass per particle, {\it k}
 the Boltzmann constant
and $v_{\mbox{s}}$ the shock velocity expressed in $\mbox{cm~sec}^{-1}$.
A formula which is useful for the implementation in the code is easily
derived, 
\begin{equation}
T = 31.80 ~ v_{\mbox{sk}}^2 ~~~{\mbox{K}}^{\circ}
,
\label{formulat}
\end{equation}
when $v_{\mbox{sk}}$ is  expressed in $\mbox{ km~sec}^{-1}$. 

\subsection {Other Effects}

Two other   effects 
are now analysed: the galactic rotation and the influence 
of the ambient pressure on  the late phase of motion.

\subsubsection{Galactic Rotation}
\label{galrot}
The effects of galactic rotation are important in shaping the
bubbles, as can be seen in figure  2 of~\citet{silich96}. 
This is very 
important on the galactic plane, and the properties of the distorted "walls"
for a rotating galaxy were described by~\citet{palous1990}.
The application of this distortion to  Gould's Belt was done by 
\citet{moreno}
, and the impact  on  large scales in a percolation-type 
scheme (in this case the cloud and star-formation mechanism is
deterministic, not stochastic) was studied by~\citet{palous94}.
The distortion  of the superbubble due to  galactic rotation is 
now introduced. The rotation curve can be that 
given by~\citet{wouterloot},
\begin{equation}
V_{\mathrm{R}} (R_0)  =220 ( \frac {R_0[\mathrm{pc}]} {8500})^{0.382} 
\mathrm {km~sec}^{-1}
,
\label {vrotation}
\end {equation} 
where $R_0$ is the radial distance from the center of the Galaxy 
expressed in pc.
The translation of the previous formula to 
the astrophysical units  adopted gives 
\begin{equation}
V_{\mathrm{R}} (R_0)  =2244 \left ( \frac {R_0[\mathrm{pc}]} 
{8500}\right )^{0.382} \frac {\mathrm{pc}}{10^7~{\mathrm{yr}}}
,
\label {vrotationpc}
\end {equation}
\begin{equation}
\Omega  (R_0)  =2244 \frac {( \frac {R_0[\mathrm{pc}]} {8500})^{0.382}} 
{R_0[\mathrm{pc}]}  \frac {\mathrm{rad}}{10^7~{\mathrm{yr}}}
.
\label {omegadiff}
\end {equation}
Here, $ \Omega (R_0)  $ is the  differential angular velocity 
and  
\begin{equation}
\phi   =2244 \frac {( \frac {R_0[\mathrm{pc}]} {8500})^{0.382}}
{R_0[\mathrm{pc}]}    t_7    ~~~\mathrm{rad}   
 .
\label {phitotale}
\end {equation}
Here, $  \phi   $, is the angle 
made on the circle and $t_7$ the time expressed
in $10^7~{\mathrm{yr}}$ units. 
Upon  considering only a  single object,
  an  expression for the
angle  can be found  once  $R=R_0+y$ is introduced, 
\begin{equation}
\phi(y)  =  \frac  {V_{\mathrm {R}} (R_0)}{R_0+y}  t 
 . 
\label{fi1} 
\end {equation}
The shift in the angle due to  differential rotation
can now be introduced,
\begin{equation}
\Delta \phi   = \phi(y) - \phi(0) 
,
\label{fi2}
\end {equation}
where {\it x } and {\it y} are the superbubble coordinates in the 
inertial frame 
of  the explosion, denoted by  {\it x}=0,{\it y}=0 and 
{\it z}=$z_{\mathrm{OB}}$.
The great distance from the center allows us to say
that   
\begin{equation}
\frac {\Delta x}{R_0} 
=   \Delta \phi 
,
\label {deltax}
\end{equation}
where  $\Delta x $ is the shift due to the differential rotation in
the {\it x} coordinate.
The shift  
can be found from (\ref{deltax}) and (\ref{fi2})
once  a Taylor  expansion is performed,  
\begin{equation}
\Delta x  \approx -  V_{\mathrm {R}} (R_0)\frac {y}{R_0} t
 .
\label {xt}
\end{equation}
Upon inserting (\ref{vrotationpc}) in (\ref{xt}),
 the following 
transformation, $T_{\mathrm {r}}$,   due to the rotation is
obtained for a single object in the solar surroundings:
\begin{equation}
T_{\mathrm{r}} ~
 \left\{ 
  \begin {array}{l} 
  x\prime=x  +  0.264 y ~t\\\noalign{\medskip}
  y\prime=y\\\noalign{\medskip}
  z\prime=z,
  \end {array} 
  \right. 
\label{trotation}
\end{equation}
where {\it y} is expressed in pc and t in $10^7~{\mathrm{yr}}$ units.
A typical  distortion introduced on a circular section of the superbubble, 
$z=z_{\mathrm{OB}}$,
is  shown in figure~\ref{rotation_worm}.

\subsection{Late Evolutionary  State}

The problem of the late evolutionary stages can be attached
including the ambient pressure; but firstly the Euler number (ratio  
of the pressure forces to the inertial forces in a flow)
 should be
introduced
\begin{equation}
Eu  = \frac  {p} { \rho_{\mathrm {s}} v_{\mathrm {s}}^2}
, 
\end{equation}
where $\rho_{\mathrm {s}}$ is the density behind the shock,
      $v_{\mathrm {s}}$    is the shock velocity and {\it p}
 the ambient pressure.
The nominal value of  pressure at the Solar
circle, $10^{-12} {\mathrm {dyne~cm}}^{-2}$,
can be used as an adequate value (see~\cite{boulares}).
Another
density distribution along the z-axis and a discussion of the pressure
distribution  can be found in ~\citet{franco}.
In order to simulate the influence of the ambient pressure,
a drag  acceleration should  be introduced,
\begin{equation}
\frac  {dy_{2,j}}{dt}    =
-v_{\mathrm {d}} ~Eu
,
\label {eq:twonumdrag}
\end {equation}
where is  $v_{\mathrm {d}}$ is an artificial  coefficient. This effect  is
negligible when the Euler number is small (high shock velocities),
and gives a  negative contribution when the Euler number $\approx$
1. 
Only in  subsection~\ref{sec464} 
are some results adopting the value
of  $v_{\mathrm {d}}=6000 {\mathrm {pc}}/10^7 ~{\mathrm{yr}}~
\approx 600 {\mathrm {km~sec}}^{-1}$  reported in
comparison with the normal run, the associated note "with the Euler
process" being always inserted.

\section {Test Calculation}

The shape   of the superbubble
depends   strongly   on
the time
 elapsed since the \underline{first} explosion, 
$t_{\mathrm {age}}$,
the \underline{duration of SN burst},
$t^{\mathrm{burst}}$,
 on
the \underline{number}
of SN explosions
in the bursting \underline{phase},
 and on the adopted density profile.

In this section, where the initial tests are discussed 
the galactic rotation is neglected.

Ordinarily, the dynamics of a superbubble is studied
with hydro simulations
(partial differential equations).
 Conversely, here under the hypothesis
called ``thin-shell approximation'', the dynamics is studied
by solving ordinary differential equations.
This choice  allows the calculation
of  a much larger number of models compared
with hydro-dynamics  calculations.

\subsection {Reliability of the code}

\label{sec:hydro}
The level  of confidence in   our results
can  be given by a comparison with numerical hydro-dynamics 
calculations  see, for example, \citet{mac}.
The vertical density distribution
they  adopted $[$ see equation(1) from~\citet{mac} and
equation(5) from~\citet{tomisaka} $]$
has  the following
{it z} dependence, 
the  distance  from  the galactic plane in pc:
\begin {equation}
n_{\mathrm {hydro}} =
n_{\mathrm {d}} \left \{
\Theta {\mathrm {exp}} [-\frac {V(z)}{\sigma_{{\mathrm {IC}}}^2} ]
+(1-\Theta) {\mathrm {exp}}
\left  [-\frac {V(z)}{\sigma_{{\mathrm {{\mathrm {C}}}}}^2} \right ]
\right\}
,
\label{ISMhydro}
\end {equation}
with the gravitational potential as
\begin{equation}
V(z)=
68.6 {\mathrm {ln}} \left 
[1 + 0.9565~{\mathrm { sinh}}^2 \left (0.758 \frac {z}{z_0}\right) \right ]
({\mathrm {km s}}^{-1})^2
.
\end {equation}
Here $n{\mathrm {_d}}=1~\mbox{particles~cm}^{-3} $, $\Theta=0.22$,
$\sigma_{IC}=14.4{\mathrm { km~s}}^{-1}$,
$\sigma_{C}=7.1  {\mathrm { km~s}}^{-1}$
and $z_0 =124~{\mathrm {pc}}$.
 Table~\ref{tab:rel} reports the results of ZEUS
(see~\cite{mac}),  a two-dimensional hydrodynamic   code, when
$t_{\mathrm {age}}=0.45 \cdot 10^7~{\mathrm{yr}}$.  
The supernova luminosity is
$1.6 \cdot 10^{38} {\mathrm{erg~s}}^{-1}$, $z_{\mathrm{OB}}$=100~pc
and
the density distribution is given by
formula~(\ref{ISMhydro}).
The ZEUS code was  originally described by~\citet{stone}.

\begin{figure}
  \begin{center}
\FigureFile(120mm,120mm){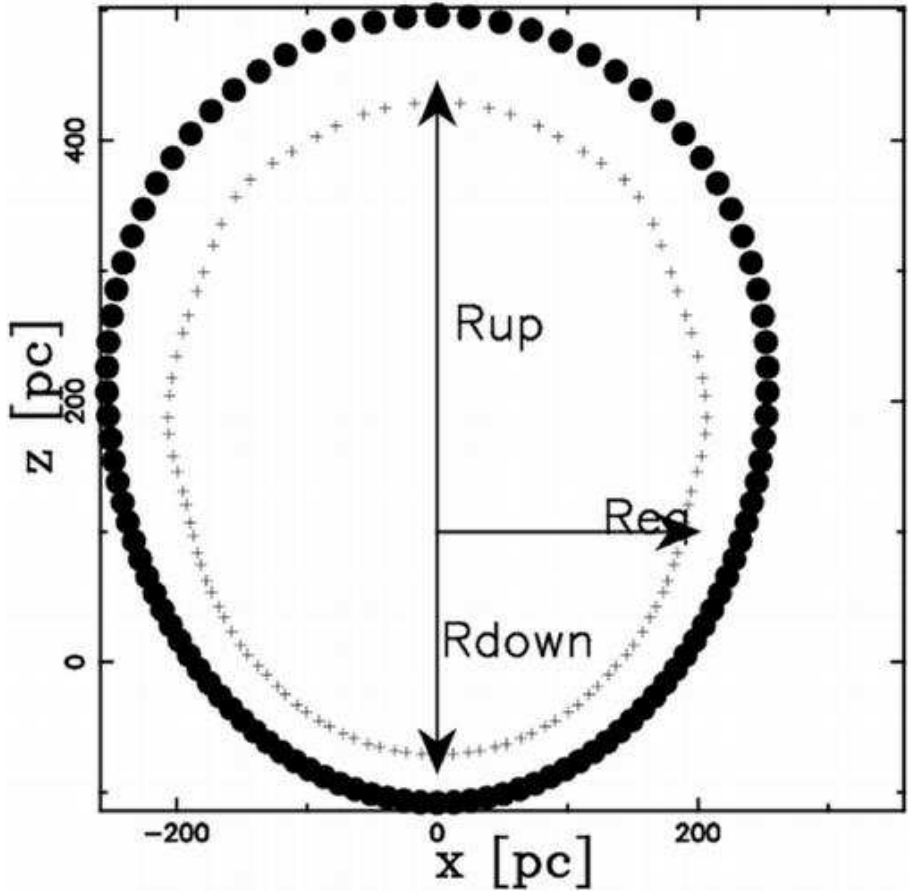}
  \end{center}
\caption
{
Section of the superbubble on the {\it x-z}  plane
when the explosion starts at  $z_{\mathrm{OB}}=100 \mbox{pc} $.
The density law is given by
equation~\ref{ISMhydro}.
The code parameters are
$t_\mathrm{\mathrm {age}}$=$0.45~10^{7}~\mathrm{yr}$,
$\Delta t=0.001\cdot10^{7}~\mathrm{yr}$,
$t^{\mathrm{burst}}_7$=0.45,
  and  $N^*$=248.
The points represented by the  small crosses indicate 
the inner section from  figure 3a  of~\citet{mac}.
}
\label{confronto}%
    \end{figure}

In order to make a comparison  our code  was run
with the parameters of the hydro-code
(see figure~\ref{confronto});
a density profile as  given by
formula~(\ref{ISMhydro}) was adopted.

From a practical point  of view,  the percentage  of
reliability  of our code can also be  introduced,
\begin{equation}
\epsilon  =(1- \frac{\vert( R_{\mathrm {hydro}}- R_{\mathrm {num}}) \vert}
{R_{\mathrm {hydro}}}) \cdot 100
\,,
\label{efficiency}
\end{equation}
where $R_{\mathrm {hydro}}$ is the   radius,
as given by the hydro-dynamics,
and  $R_{\mathrm {num}}$ the radius  obtained from our  simulation.
In the already cited table~\ref{tab:rel},
our numerical radii can also be found
in the    upward, downward and  equatorial
directions,
  and the efficiency  as  given by
formula~(\ref{efficiency}).
The value of the radii are comparable  in
all of the three  chosen directions.
figure~\ref{confronto} also reports on   the data
from  figure 3a  of~\citet{mac}.

\begin{table}
      \caption{Code reliability. }
         \label{tab:rel}
      \[
         \begin{array}{ccccc}
            \hline
            \noalign{\smallskip}
~~~~     & R_{\mathrm{up}}(\mathrm{pc})  &
         R_{\mathrm{down}}(\mathrm{pc})  &
         R_{\mathrm{eq}}  (\mathrm{pc})  \\
            \noalign{\smallskip}
            \hline
            \noalign{\smallskip}
R_{\mathrm {hydro}} (ZEUS)     &  330  &  176   &  198        \\
R_{\mathrm {num}}   (\mbox{our~code}) &  395  &  207  &  237
        \\
\mbox {efficiency} (\%)      &  80  &  81  &   80        \\
            \noalign{\smallskip}
            \hline
         \end{array}
      \]
   \end{table}

\citet{koo0} applied their
one-dimensional method based on the virial
theorem to a two-dimensional calculation by using a
sector approximation,
which is identical to that assumed in
the present paper and in which a
differential equation is solved along a radial path.
figure 8 from~\citep{koo0}  is for the same
density distribution
as given by  formula~(\ref{ISMhydro}).

\subsection{The Asymmetries}
\label{sec_asym}

\begin{figure}
\FigureFile(120mm,120mm){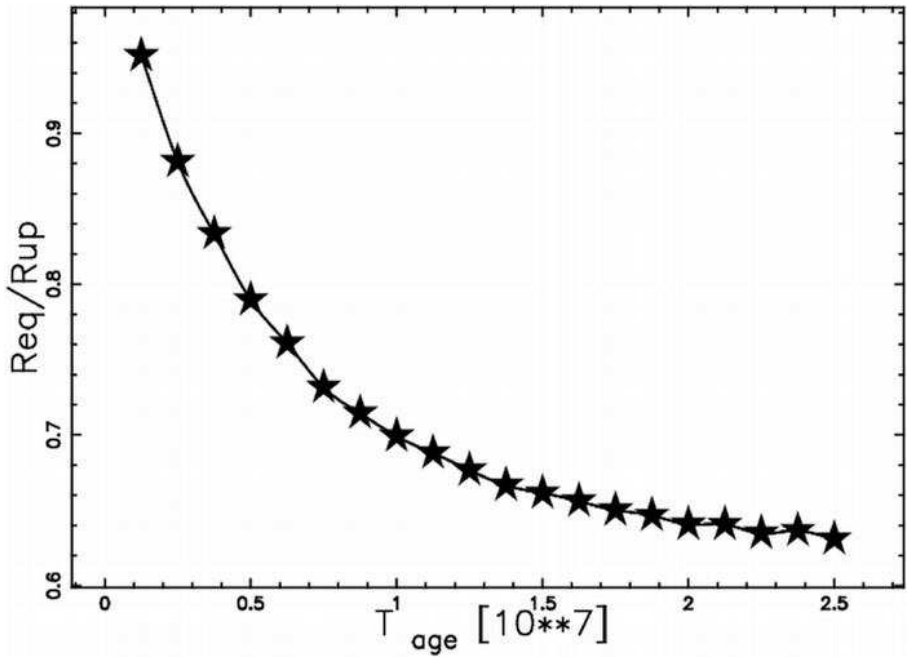}
\caption
{$R_{\mathrm{eq}}/R_{\mathrm{up}}$  versus  the  time,
$t_{\mathrm {age}}$,   expressed in
units of $10^{7}~\mathrm{yr}$~
when the explosion starts at $z_{\mathrm{OB}}=0 \mbox{pc}$.
The parameters are
$\Delta t=0.1 \cdot 10^{7}~\mathrm{yr}$,
$t^{\mathrm{burst}}_7$= 0.45,  and  $N^*$=  248.
}
\label{figreqrup}%
    \end{figure}

Firstly,  the time evolution of
the ratio $R_{\mathrm{eq}}/R_{\mathrm{up}}$ is focused on as
a function of $t_{\mathrm {age}}$.
figure~\ref{figreqrup}~ plots the asymmetry indicator,
$R_{\mathrm{eq}}/R_{\mathrm{up}}$ for a  model
with $z_{\rm OB}=0$, $t^{\mathrm{burst}}_7=0.45$
and $N^*=250$.
In this case,  the indicator  of the asymmetry,  $R_{\mathrm{eq}}/R_{\mathrm{up}}$
is  $\approx$ 0.9 at $t_{\mathrm {age}}$= $0.4 \cdot 10^{7}~\mathrm{yr}$,
reaches a minimum at $t_{\mathrm {age}}$= $2.0 \cdot 10^{7}~\mathrm{yr}$
and remains nearly constant up  to
$t_{\mathrm {age}}$= $2.5 \cdot 10^{7}~\mathrm{yr}$~
(see figure~\ref{figreqrup}~).

   %
\begin{figure}
\FigureFile(120mm,120mm){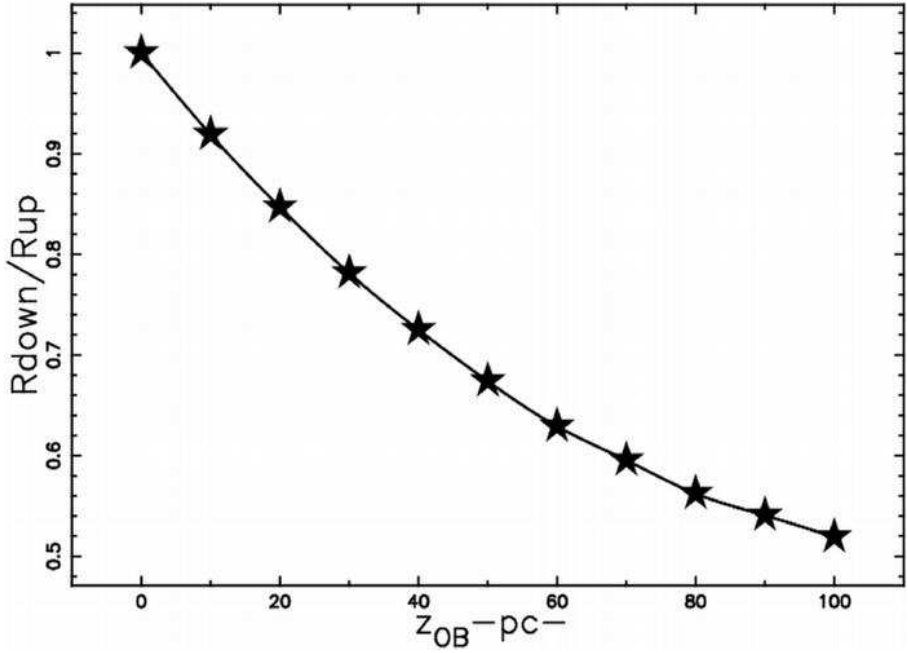}
\caption
{$R_{\mathrm{down}}$/$R_{\mathrm{up}}$  versus  $z_{\mathrm{OB}}$ in pc.
The parameters are
$t_\mathrm{\mathrm {age}}$=$0.45 \cdot 10^{7}~\mathrm{yr}$,
$\Delta t=0.005 \cdot 10^{7}~\mathrm{yr}$,
$t^{\mathrm{burst}}_7$= 0.45,  and  $N^*$=  248.
}
\label{figrdownz}%
    \end{figure}

Secondly, we consider  how the asymmetry indices,
$R_{\mathrm{down}}/R_{\mathrm{up}}$ (figure~\ref{figrdownz})
and $R_{\mathrm{eq}}/R_{\mathrm{up}}$ (figure~\ref{figrupz}), depend
upon the height of the OB association, $z_{\rm OB}$.
A comparison is made at $t_{\rm age}=4.5\cdot 10^6$yr
and $N^*=248$.
Now,  the first index  of the asymmetry, $R_{\mathrm{down}}/R_{\mathrm{up}}$,
is 1  at $z_{\rm OB}$=0~pc  and 0.52 at  $z_{\rm OB}$=100~pc
(value of $z_{\rm OB}$ where the comparison with the hydro-code
 was performed; see figure~\ref{figrdownz}).
Another interesting theoretical parameter that has an
observational counterpart is the degree of symmetry with  respect to
the galactic plane. Upon  adopting the previously developed notation, 
 two variables, $D_{\mathrm{up}}$ and  $D_{\mathrm{down}}$,
 are
introduced when $R_{\mathrm{down}}~>~z_{\mathrm{OB}}$:
\begin{eqnarray}
R_{\mathrm{up}}  +z_{\mathrm{OB}}=& D_{\mathrm{up}}\nonumber,   \\
R_{\mathrm{down}} -z_{\mathrm{OB}}=& D_{\mathrm{down}}  . \label{defD}
\end {eqnarray}

\begin{figure}
\FigureFile(120mm,120mm){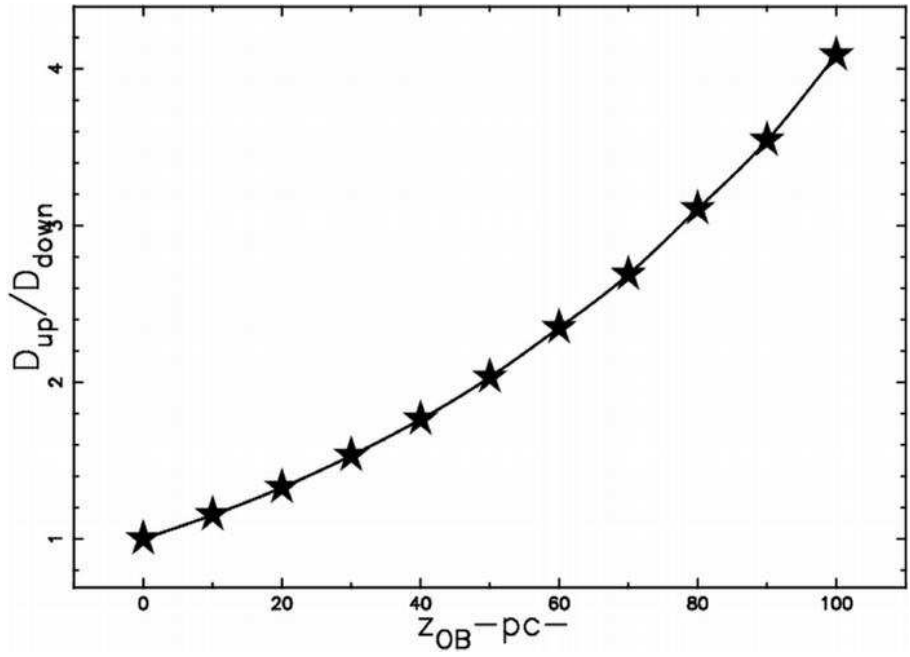}
\caption
{
Ratio $\frac{ D_{\mathrm{up}}} { D_{\mathrm{down}}}$
versus  $z_{\mathrm{OB}}$ in pc.
The parameters are the same as in  figure~\ref{figrdownz}.
}
\label{figrdownz_lat}%
    \end{figure}

\begin{figure}
\FigureFile(120mm,120mm){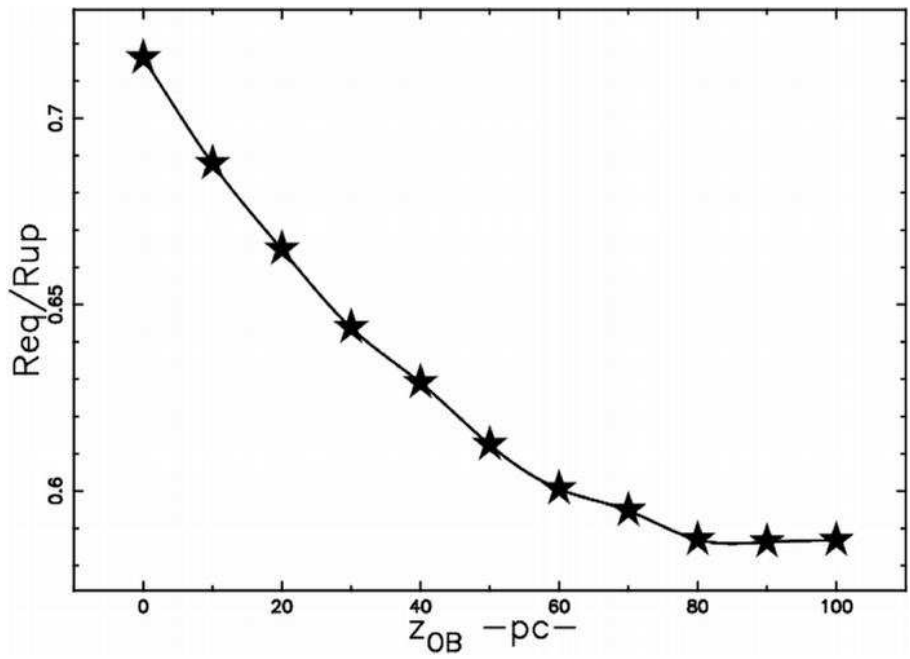}
\caption
{
$R_{\mathrm{eq}}$/$R_{\mathrm{up}}$  versus  $z_{\mathrm{OB}}$ in pc.
The physical  parameters are the same
as  in figure~\ref{figrdownz}.
}
\label{figrupz}%
    \end{figure}

The super-shells are often visualised in galactic coordinates.
When this  visualisation is possible,  the ratio 
between  the positive 
maximum galactic latitude    and                                                            
the absolute value of the negative minimum 
 galactic latitude
 belonging to a given super-shell
can be identified with $\frac{ D_{\mathrm{up}}} { D_{\mathrm{down}}}$.
In our theoretical framework, 
the ratio  is easily found
to be 
\begin{equation}
\frac{ D_{\mathrm{up}}} { D_{\mathrm{down}}}= \frac {R_{\mathrm{up}}+ z_{\rm OB}}
                                            {R_{\mathrm{down}} - z_{\rm OB}}
 ,
\label{ratio_lat}
\end {equation}
and  can be visualised in 
figure~\ref{figrdownz_lat};
due to the nature of the phenomena,
 this ratio easily takes high values.

Conversely, the  second  index,
$R_{\mathrm{eq}}$/$R_{\mathrm{up}}$,
as  a function  of $z_{\mathrm{OB}}$
presents a marked minimum value at $\approx$ 100~pc;
for this value of $z_{\mathrm{OB}}$
the asymmetry in the shape of
the expanding envelope is maximum
(see figure~\ref{figrupz}).
   %

   %

\subsection{Characteristic Shapes}
\label{shapes}

The  developed  theory  of bubble expansion
in the ISM   has  four key parameters: the distance  $z_{\mathrm{OB}}$
from the galactic plane , $t^{\mathrm{burst}}_7$,
  the bursting \underline{phase}, $N^*$,   the number of SN explosions
in  $2.5 \cdot 10^7$ \mbox{yr}, and $t_{\mathrm {age}}$ the
age of the supershell . By varying  these four
parameters,  a great variety of different shapes can be
obtained and  two   cases, $z_{\mathrm{OB}}$=0~pc
and $z_{\mathrm{OB}}$=100~pc,
will be concentrated on.

\subsubsection {From the ellipse to the hourglass, $z_{\mathrm{OB}}$=0~pc}

\begin{figure}
\FigureFile(120mm,120mm){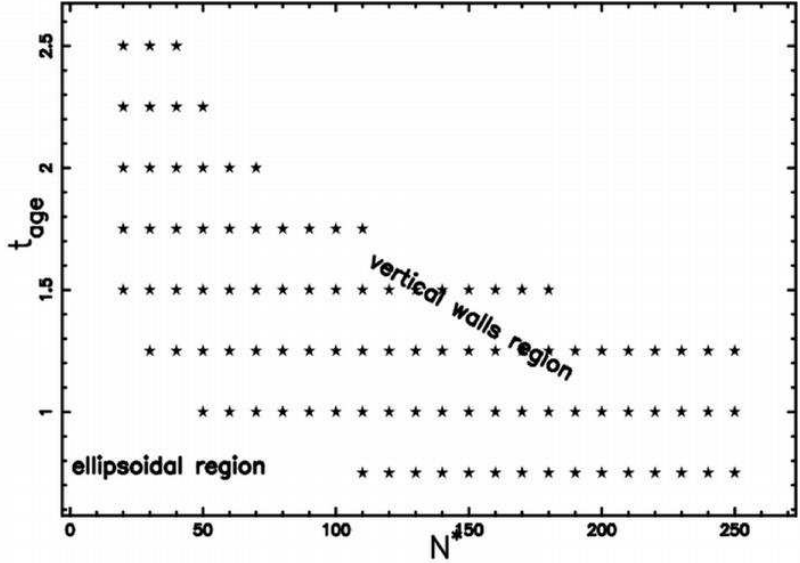}
\caption
{
Stars indicate   the existence of 
"vertical walls"; $t_{\mathrm {age}}$ is expressed in $10^7$yr units.
The parameters are
$\Delta t=0.1 \cdot 10^{7}~\mathrm{yr}$, $z_{\mathrm{OB}}$=0, and
$t^{\mathrm{burst}}=t_{\mathrm {age}}$.
}
\label{conditions}%
    \end{figure}

\begin{figure}
\FigureFile(120mm,120mm){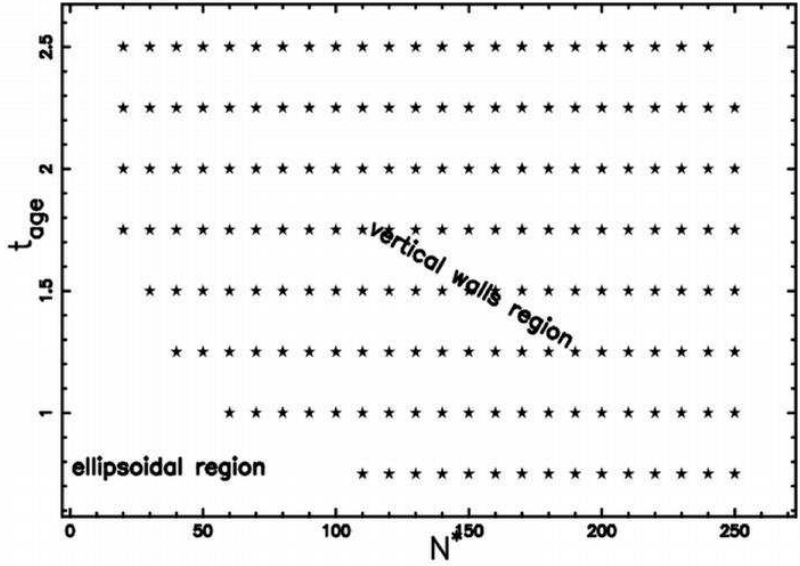}
\caption
{
Stars indicate   the existence of the
"vertical walls"; $t_{\mathrm {age}}$ is expressed in $10^7$yr units.
The parameters are
$\Delta t=0.1 \cdot 10^{7}~\mathrm{yr}$, $z_{\mathrm{OB}}$=0\mbox{pc}, and
$t^{\mathrm{burst}}_7=0.5$.
}
\label{conditions2}%
    \end{figure}

\begin{figure}
\FigureFile(120mm,120mm){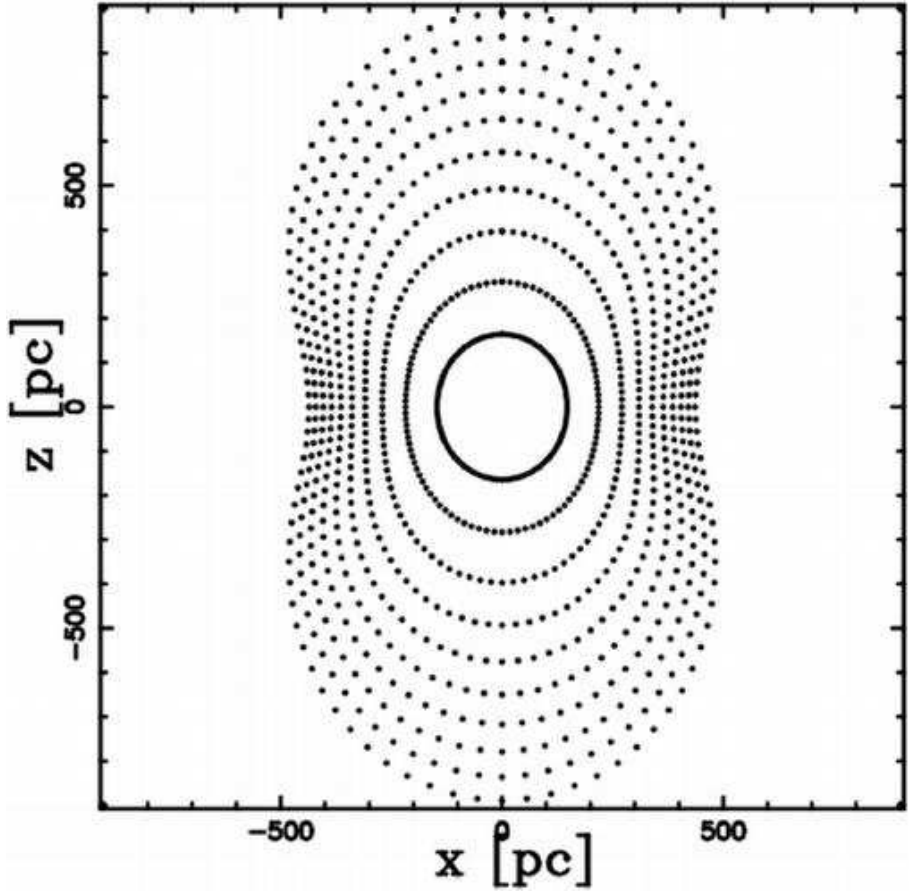}
\caption
{Section of the superbubble on  the {\it x-z}  plane.
The parameters are
$t_\mathrm{\mathrm {age}}$=$2.5 \cdot 10^{7}~\mathrm{yr}$,
$\Delta t=0.01 \cdot 10^{7}~{\mathrm{yr}}$, $z_{\mathrm{OB}}$=0\mbox{pc},
$t^{\mathrm{burst}}_7$= 0.5,  and  $N^*$=  100.
The sections at regular  intervals of  $t_{\mathrm {age}}$/10
are also shown.
}
\label{vertical}%
    \end{figure}

We start by fixing $z_{\mathrm{OB}}$=0~pc (the  symmetrical case in the
+{\it z} and -{\it z} directions).
If the luminosity is high enough and $z_{\mathrm{OB}}=0$,
the superbubble changes its shape
from  spheroidal to hourglass shape via the vertical wall.
In order  to define  the vertical wall,
a reasonable angle of $30^{\circ}$
is first introduced
 between the plane
$z=0$ and a certain radius, $R_{30^{\circ}}$ ; this radius
projected on the {\it z}=0 plane is  denoted by
$R_{p,30^{\circ}}$.
From this point of view,
a vertical
wall is made if $R_{z=0}\simeq R_{p, 30^{\circ}}$;
$R_{z=0} >  R_{p , 30^{\circ}}$ indicates the elliptical shape and
$R_{z=0} <  R_{p , 30^{\circ}}$ indicates the hourglass  shape.
Another way  to express the previous definition
is the following inequality, which  can
be easily
embedded in the code:
\begin {equation}
\frac  { | R_{\mathrm{eq}} - R_{p,30^{\circ}}| } { R_{\mathrm{eq}}}  \cdot 100
\leq 5 \%
 .
\end {equation}
Through  such a definition,  the vertical walls will span $60^{\circ}$
on the galactic height.
The existence of the vertical wall  can now be plotted
in the $t_{\mathrm {age}}-N^*$ space;
see  figure~\ref{conditions} when $t^{\mathrm{burst}}$= $t_{\mathrm {age}}$,
and  figure~\ref{conditions2} when $t^{\mathrm{burst}}_7$=0.5.
\label{verticalwalls}
From these two plots (\ref{conditions} and~\ref{conditions2})
it  can be inferred  that:
\begin {itemize}
\item   When   $t^{\mathrm{burst}}_7$= 0.5,
        vertical walls are only  developed  if $N^*\ge~20$;
        for example  at $N^*$=100
        they appear at  $t_{\mathrm {age}}~\approx~1.0 \cdot 10^7$ yr
        (see figure \ref{conditions2}).
        In this case  the transition
        from an  ellipse to an  hourglass
         is not observed;  in other words, the vertical walls
         are the typical structure over a wide range of
         $t_{\mathrm {age}}$ and $N^*$.
\item   When   $t^{\mathrm{burst}}$= $t_{\mathrm {age}}$
        the transition 
         between the various zones is better defined;
         for example,  at $N^*$=50
        the vertical walls  appear at
        $t_{\mathrm {age}}~\approx~1.0 \cdot 10^7$ yr (see figure \ref{conditions}).
\end {itemize}
A typical  plot showing the temporal  development  of
  vertical  walls  through a displacement of the sections
at  regular time intervals
is reported  in figure~\ref{vertical}.
Further on,  from
the data  of plots (\ref{conditions} and \ref{conditions2}),
we can extract an approximate age,  after which  the vertical
walls appear when $t^{\mathrm{burst}}_7$= 0.5, $t^{0.5} _{\mathrm{VW}}$
and  when  $t^{\mathrm{burst}}$= $t_{\mathrm {age}}$, $t^{\mathrm {age}} _{\mathrm{VW}}$.
The numerical  law can  be obtained by fitting
the  value $t_{\mathrm{VW}}$ (the value of $t_{\mathrm {age}}$
at which the vertical walls appear) versus $N^*$,
\begin {equation}
\log (t_{\mathrm{VW}}) =  a +b \cdot  \log (N^*)
 .
\end {equation}
The values of a and b  are   easily  derived from the least-squares
fitting procedure.
In other words,
 this relationship connects  $N^*$ and the first
corresponding  value of  $t_{\mathrm {age}}$ at the left
of figure~\ref{conditions}, and figure~\ref{conditions2}
is defined as  $t_{\mathrm{VW}}$.
The  application of  least-squares   gives
\begin{equation}
t^{0.5}_{\mathrm{VW}} =  4.5  (N^*)^{-0.34} \cdot  10^7  {\mathrm{yr}}
\label{vw1}
 ,
\end {equation}
\begin{equation}
t^{\mathrm {age}}_{\mathrm{VW}} =  3.2 (N^*)^{-0.28} \cdot 10^7  {\mathrm{yr}}
 .
\label{vw2}
\end {equation}
It can  therefore  be concluded  that   introducing 
of a finite  value of  $t^{\mathrm{burst}}$ (for example
$t^{\mathrm{burst}}_7$= 0.5)  inhibits the transition from
the vertical walls to the hourglass shape.
It is important to note  that the time necessary to develop 
vertical  walls decreases when the key parameter, $N^*$,
increases; the exact law being an inverse  power.
In other words, the momentaneous
radius of the evolving superbubble in the galactic plane
({\it z}=0) gives a  reference radius; the balance between a  
progressively  greater distance at higher values of {\it z}
and the decreasing matter encountered along the trajectory
allows the radius along the other directions
to be greater with respect to that along the galactic plane.
Under the existence of certain parameters,  as given
by  the numerical laws~(\ref{vw1}) and ~(\ref{vw2}),
a  progressively  increasing radius (connected with the
progressive angle with respect to the galactic plane)
 having a  "vertical shape"  is obtained.

\subsubsection{ Toward the V-shape, $z_{\mathrm{OB}}>0~pc$}

\label{sec:egg}

\begin{figure}
\FigureFile(120mm,120mm){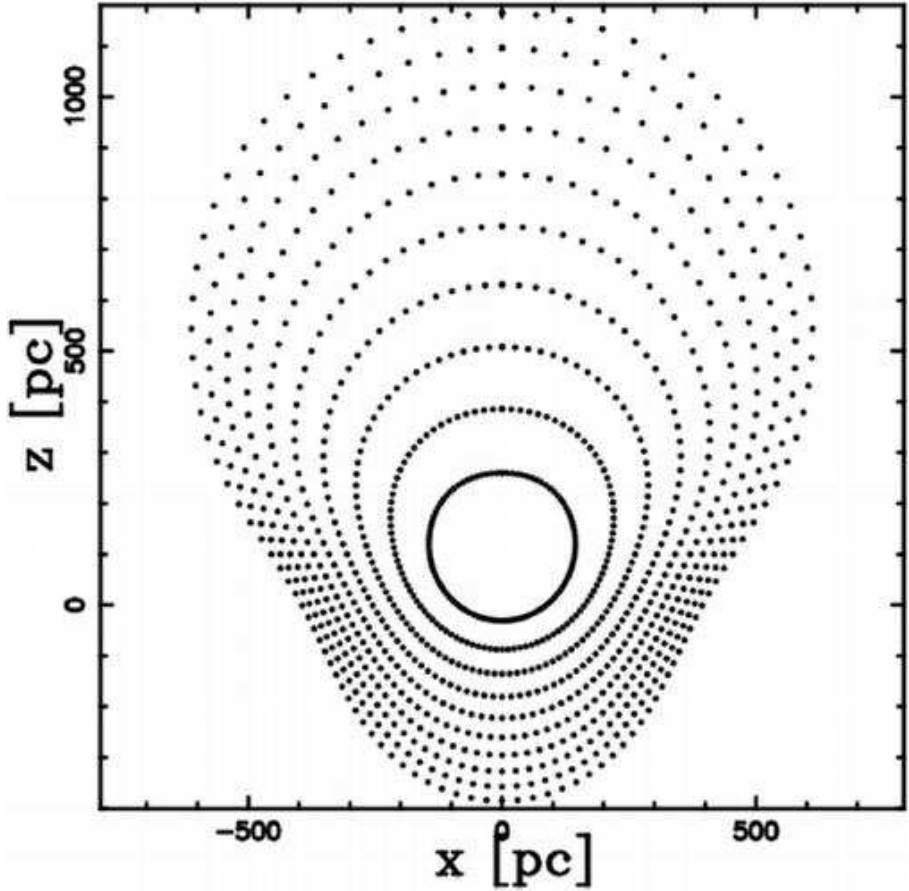}
\caption
{
Section of the superbubble on the {\it x-z } plane.
The parameters are
$t_\mathrm{\mathrm {age}}$=$2.5 \cdot 10^{7}~\mathrm{yr}$,
$\Delta t=0.01 \cdot 10^{7}~\mathrm{yr}$, $z_{\mathrm{OB}}$=100~\mbox{pc},
$t^{\mathrm{burst}}_7$= 0.5,   and  $N^*$=  250.
The sections at regular  intervals of  $t_{\mathrm {age}}$/10
are also shown.
}
\label{vshape}%
    \end{figure}

A substantial modification of the up/down symmetry  with
respect to the galactic plane can be obtained
by varying the distance
$z_{\mathrm{OB}}$ from the galactic plane.
In this case
the morphologies become complicated and we limit ourselves
to exploring   the situation at $z_{\mathrm{OB}}$=100~pc where
a transition from the egg  shape to the
V-shape is observed (see
figure~\ref{vshape});
in this case the
temporal evolution of the superbubble is  visualised
through a displacement of the sections
at  regular time intervals.

When $z_{\mathrm{OB}}$=100~pc the  density in the equatorial plane
of the explosions is  $\approx$ two-times lower with
respect
to the density of  the galactic
plane. The surface of the expansion will therefore find an
increasing  density in the downward direction and, conversely,
a decreasing density in the
upward direction; the curious
V-shape is generated from the competition of these two effects.

\begin{figure}
\FigureFile(120mm,120mm){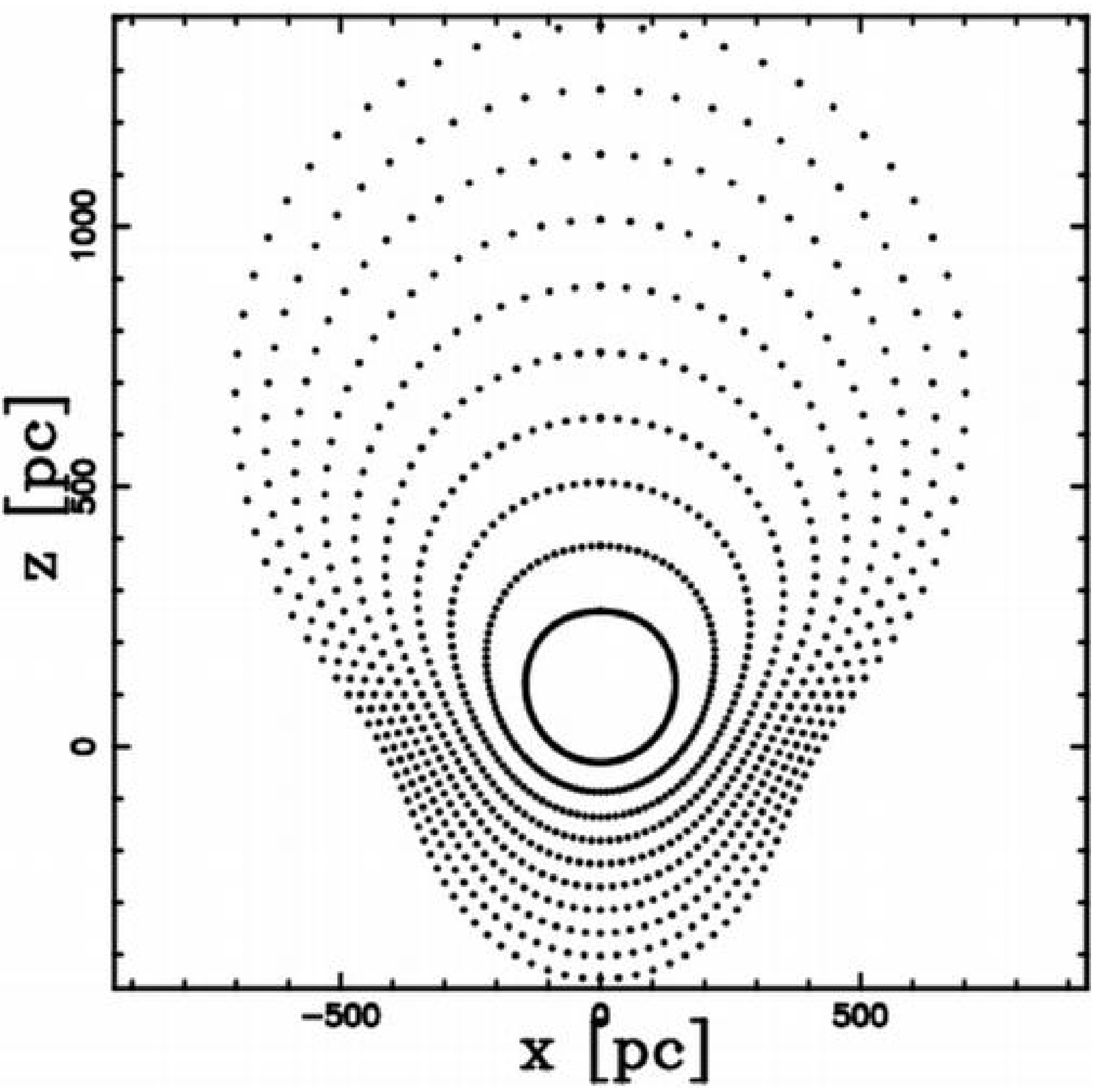}
\caption
{
Same as figure~\ref{vshape},  but with
$t^{\mathrm{burst}}$= $t_{\mathrm {age}}$.
}
\label{vshape_burst}%
    \end{figure}

The effect  of increasing   the bursting time  is 
shown in figure~\ref{vshape_burst},
where $t^{\mathrm{burst}}$= $t_{\mathrm {age}}$ rather than 
 $t^{\mathrm{burst}}=0.5 \cdot 10^7~{\mathrm{yr}}$ of  figure~\ref{vshape}.
In the case  of  figure~\ref{vshape_burst}, 
the maximum elongation along the {\it z}-direction is  $\approx$ 1400 pc
against  $\approx$ 1100 pc
of  figure~\ref{vshape}, 
in which a limited time of bursting was chosen.

\section{Single Astrophysical Objects}

Another  way of detecting   supershells is through
a  search  for
galactic worms;  they are filamentary, vertical structures
that  look like worms crawling away  from the galactic
plane  in  H\,I and infrared   surveys
(~60  and  100$\mu m$)~\citep{koo}.

With respect to the worms, there are also several alternatives to
their origin. 
A  possibility is due to gravo-magnetic
instabilities,  such as the Parker instability in 3D, 
which forms worm-like
structures, as can be  seen in figure  4b 
of~\cite{franco2002}.

Here,  it is  assumed that they are the vertical walls of
supershells,  as  sustained by an observational
point of view ~\citep{kim}  or  by a theoretical
argument  (see  appendix~\ref{sec_ring}).

The developed theory is now applied to a well-defined
galactic supershell  (see  subsection~\ref{sec464} and
subsection~\ref{sec238} ).

In this astrophysical  section 
the galactic rotation is inserted.

\subsection{Supershell Associated with GW~46.4+5.5}
\label{sec464}

A careful study of the worms 46.4+5.5 and  39.7+5.7~\citep{kim}
has led to the conclusion that they  belong to a single
super-shell. Further on,  the dynamical properties  of this H\,I 
supershell can be deduced by coupling the observations with 
theoretical arguments,~\citet{igu}; 
the derived  model parameters to fit observations 
are reported in  table~\ref{tab:ssh},
where the  altitudes of 
KK~99~3 and KK~99~4   (clouds that are CO emitters)
have been identified with $z_{\mathrm{OB}}$ by the author.

   \begin{table}
      \caption{Data of the supershell associated with GW~46.4+5.5.}
         \label{tab:ssh}
      \[
         \begin{array}{cc}
            \hline
            \noalign{\smallskip}
\mbox {Size~(pc}^2)                    & 345 \cdot 540  \\
\mbox {Expansion~velocity~ (km~s$^{-1}$}) & 15              \\
\mbox {Age~(10$^7$~yr})                    & 0.5             \\
\mbox {z$_{OB}$  (pc)}                   & 100             \\
\mbox {Total~energy~($10^{51}$}{\mathrm{erg}})      & 15              \\
            \noalign{\smallskip}
            \hline
         \end{array}
      \]
   \end{table}
 
\begin{figure}
\FigureFile(120mm,120mm){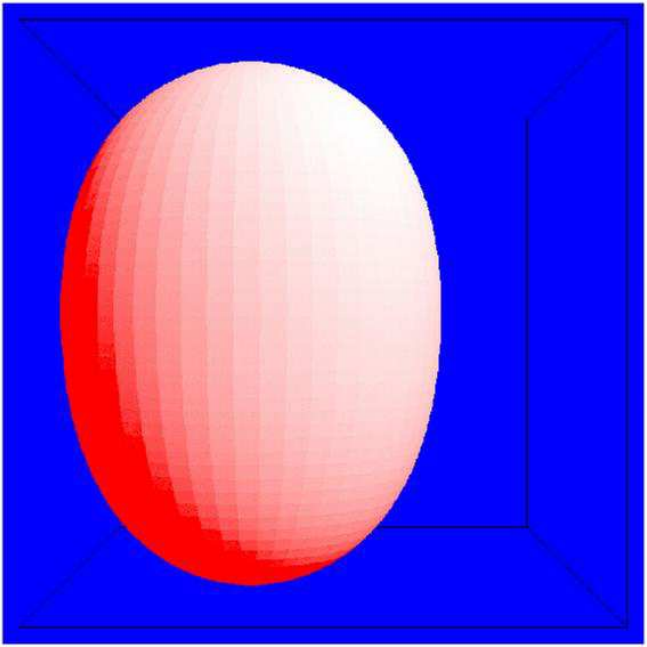}
\caption
{
Model of  GW~46.4+5.5.
The parameters are
$t_\mathrm{\mathrm {age}}$=$0.5 \cdot 10^{7}~\mathrm{yr}$,  
$\Delta t=0.001 \cdot 10^{7}~\mathrm{yr}$,
$t^{\mathrm{burst}}_7$= 0.5, $N^*$=  150, $z_{\mathrm{OB}}$=100 pc, 
and $E_{51}$=1.
The three Eulerian angles 
characterising the point of view of the observer
are  $ \Phi $= 0$^{\circ }$
, $ \Theta $= 90$^{\circ }$
 and  $ \Psi $= 0$^{\circ }$.
}
\label{worm}%
    \end{figure}

These parameters   are  the input for  our computer 
code (see the captions of figure~\ref{worm}).
The problem of assigning a value to $z_{\mathrm{OB}}$ now arises,
 and the
following two equations are set up:
\begin{eqnarray}
R_{\mathrm{up}}  +z_{\mathrm{OB}}=& 540&~  , \nonumber   \\
\frac{R_{\mathrm{up}}+z_{\mathrm{OB}}}{R_{\mathrm{down}}-z_{\mathrm{OB}}}=& \frac{D_{\mathrm{up}}}{D_{\mathrm{down}}}= &
\frac{15^{\circ}}{3^{\circ}}=3.
\label{system}
\end {eqnarray}
The algebraic system~(\ref{system}) consists in two equations and
three variables. The value chosen for the minimum and maximum
latitude ($15^{\circ}$    and   $-5^{\circ}$)
 is in rough agreement with the position of the center at
$+5^{\circ}$ of the galactic latitude (see \cite {koo}). 
One way 
of solving 
system  \ref{system} is to set, for example, $z_{\mathrm{OB}}$=100~pc.
The other two variables are easily found to be as  follows:
\begin{eqnarray}
z_{\mathrm{OB}}   =& 100~\mbox {pc}  ,\nonumber   \\
R_{\mathrm{down}} =& 235~\mbox {pc}  ,           \\
R_{\mathrm{up}}  =& 305~\mbox  {pc}
 \nonumber  .
\label{solution}
\end {eqnarray}
For this value of 
  $z_{\mathrm{OB}} $,   we 
have  the case where a transition from egg-shape to V-shape
(as  outlined in subsubsection~\ref {sec:egg}) 
is going on,    and the simulation gives the exact shape.    
In order to obtain 
$E_{\mathrm{tot}}=15. \cdot 10^{51}{\mathrm{erg}}$ 
(see formula~(\ref{etotburst}))
and  $t^{\mathrm{burst}}=0.5 \cdot 10^7~{\mathrm{yr}}$,
we have inserted 
$N^*$=~150.

In order to  test our simulation,  an
observational percentage  of 
reliability is  introduced that uses 
both the size and the shape, 
\begin{equation}
\epsilon_{\mathrm {obs}}=100(1-\frac{\sum_j |R_{\mathrm {obs}}-R_{\mathrm {num}}|_j}{\sum_j
{R_{\mathrm {obs}}}_j})
, 
\label{eq:reliability}  
\end{equation}
where $R_{\mathrm {obs}}$ is the observed  radius,
as deduced  by 
using the following algorithm. 
The radius at regular angles from the  vertical 
($0{^{\circ}}, ~90{^{\circ}}, ~180{^{\circ}}$) is extracted from
table~\ref{tab:ssh},
 giving the series (305~pc, ~172.5~pc, ~235~pc);
the cubic spline theory~\citep{press}  
is then applied to compute
the various radii at 
progressive angles 
($0{^{\circ}}, ~3.6{^{\circ}}, ~7.2{^{\circ}},...180{^{\circ}}$),
which  is a series computed by adding 
regular steps of  $180~{^{\circ}}/n_{\theta}$. 
The data are extracted from the dotted ellipse
visible in figure~7 of \cite{koo};
this  ellipse represents the super-shell at 
$v_{LSR}=18.5~\mbox{km\,sec}^{-1}$.

\begin{figure}
\FigureFile(120mm,120mm){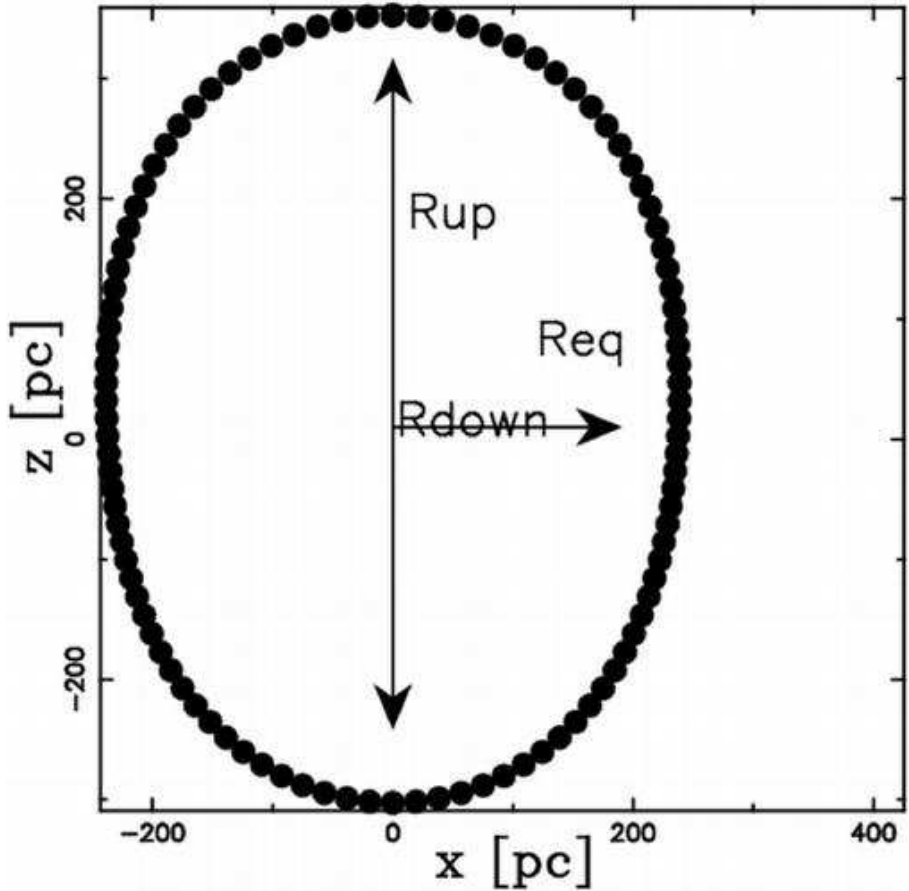}
\caption
{Section of the superbubble on the {\it x-z }
 plane    when the   
physical parameters are the  same as in figure~\ref{worm}.
}
\label{worm_sect}%
    \end{figure}

We can now compute the efficiency over 50+1 directions
$[$ formula~(\ref{eq:reliability})$]$  of a section {\it y-z} when
 {\it x}=0  
which  turns out to be 
$\epsilon_{\mathrm {obs}}=68.4 \%$  
($\epsilon_{\mathrm {obs}}=71.3 \%$ with the Euler process);
the observed and numerical
radii along  the three typical directions are reported
in table~\ref{tab:rel_gw}.
The physical parameters adopted from~\citep{kim}  
turn out to be consistent with our numerical code.
   \begin{table}
      \caption{Radii concerning  GW~46.4+5.5.}
         \label{tab:rel_gw}
      \[
         \begin{array}{cccc}
            \hline
            \noalign{\smallskip}
\mathrm{Direction}& R_{\mathrm {num}}(\mathrm{pc}) &R_{\mathrm
{num}}(\mathrm{pc})
\mbox { with
 the Euler process}& R_{\mathrm {obs}}(\mathrm{pc}) \\
            \noalign{\smallskip}
            \hline
            \noalign{\smallskip}
\mathrm{Equatorial} &  238  & 233 &   172.5   \\
\mathrm{Polar~up}   &  342  & 335 &   305     \\
\mathrm{Polar~down} &  312  & 304 &   235     \\
            \noalign{\smallskip}
            \hline
         \end{array}
      \]
   \end{table}
The result of the simulation can be visualised in~figure~\ref{worm},
or through a  section on the {\it x-z}
 plane  (see 
figure~\ref{worm_sect}).
\label{ref_velocity}
Another  important  observational  parameter is the H\,I
21~cm line emission: the observation of H\,I gas associated 
with GW~46.4+5~\citep{hartmann,kim}  
reveals that the super-shell has 
an expansion velocity 
of  $V_{\mathrm{exp}} \approx$ 15  $\mbox{km\,s}^{-1}$~\citep{kim}.
In order to see how our model matches  the observations,
the instantaneous radial velocities 
are computed in each direction.
A certain number of random points are then generated 
on the theoretical surface of expansion. The relative velocity
of each point is  computed
by using the method  of   bilinear interpolation on the four 
grid points that surround the selected latitude and 
longitude~\citep{press}. \label{para:velo}
 
\begin{figure}
\FigureFile(120mm,120mm){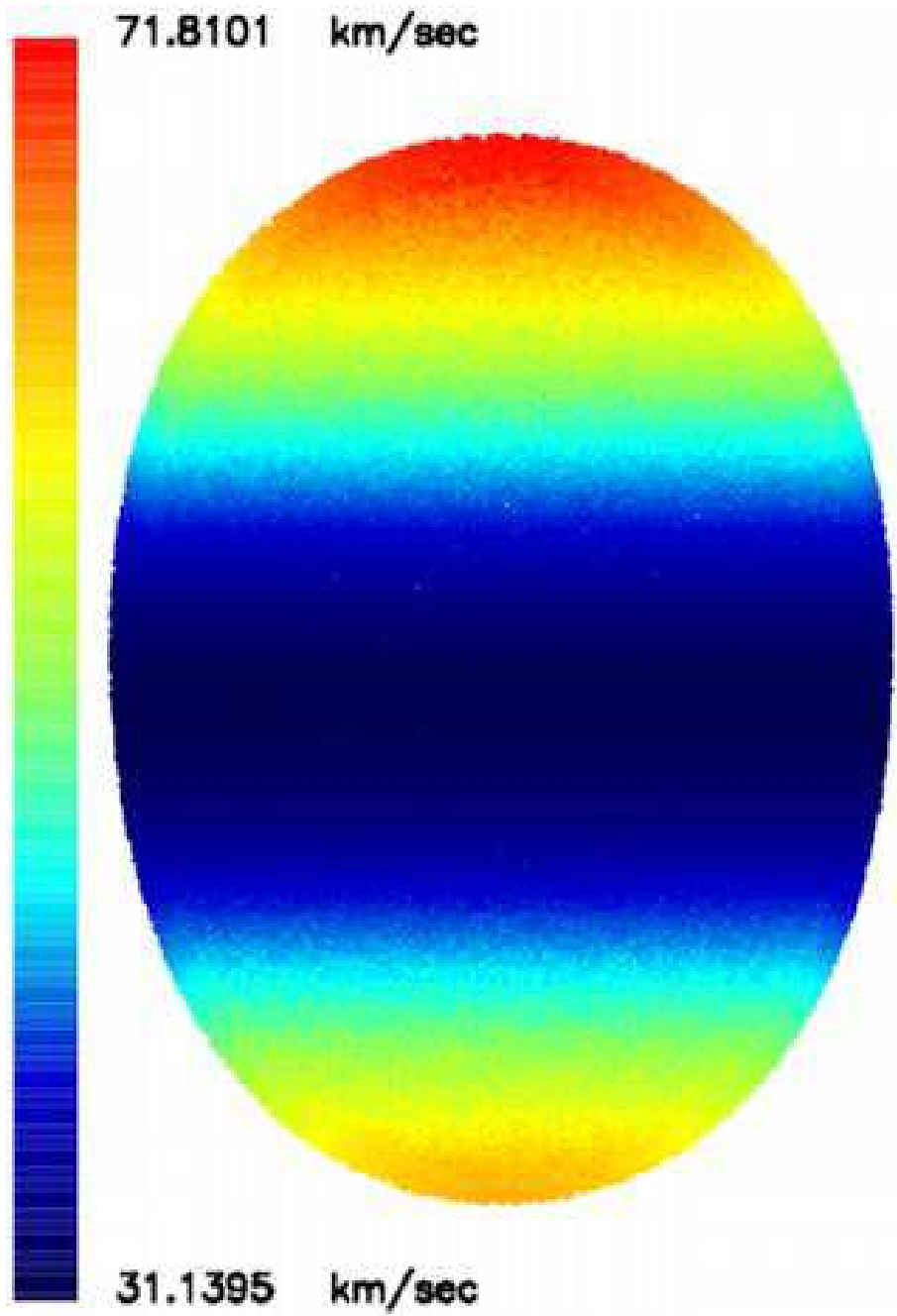}
\caption
{
  Map of the expansion velocity  
  relative to a simulation of GW~46.4+5.5~ 
  when 190000 random points are selected on the surface.
  The   physical parameters are the  same as in figure~\ref{worm}
  and 
  the three Eulerian angles 
  characterising the point of view of the observer
  are  $ \Phi $= 0   $^{\circ }$
, $ \Theta $= 90   $^{\circ }$,
  and  $ \Psi $=0   $^{\circ }$.
}
\label{worm_velocita}%
    \end{figure}

Our model gives radial velocities of, $V_{\mathrm{theo}}$, 
31~ $\mbox{km~ s}^{-1}$ $\leq~V_{\mathrm{theo}}~\leq$ 71 ~
$\mbox{km~s}^{-1}$
(27~ $\mbox{km~s}^{-1}$ $\leq~V_{\mathrm{theo}}~\leq$ 66 ~
$\mathrm{km~s}^{-1}$
with the Euler process)
and a  map of the expansion velocity 
is given in figure~\ref{worm_velocita},
from which it is possible to visualise the differences 
in the expansion velocities among the various regions.

\begin{figure}
\FigureFile(120mm,120mm){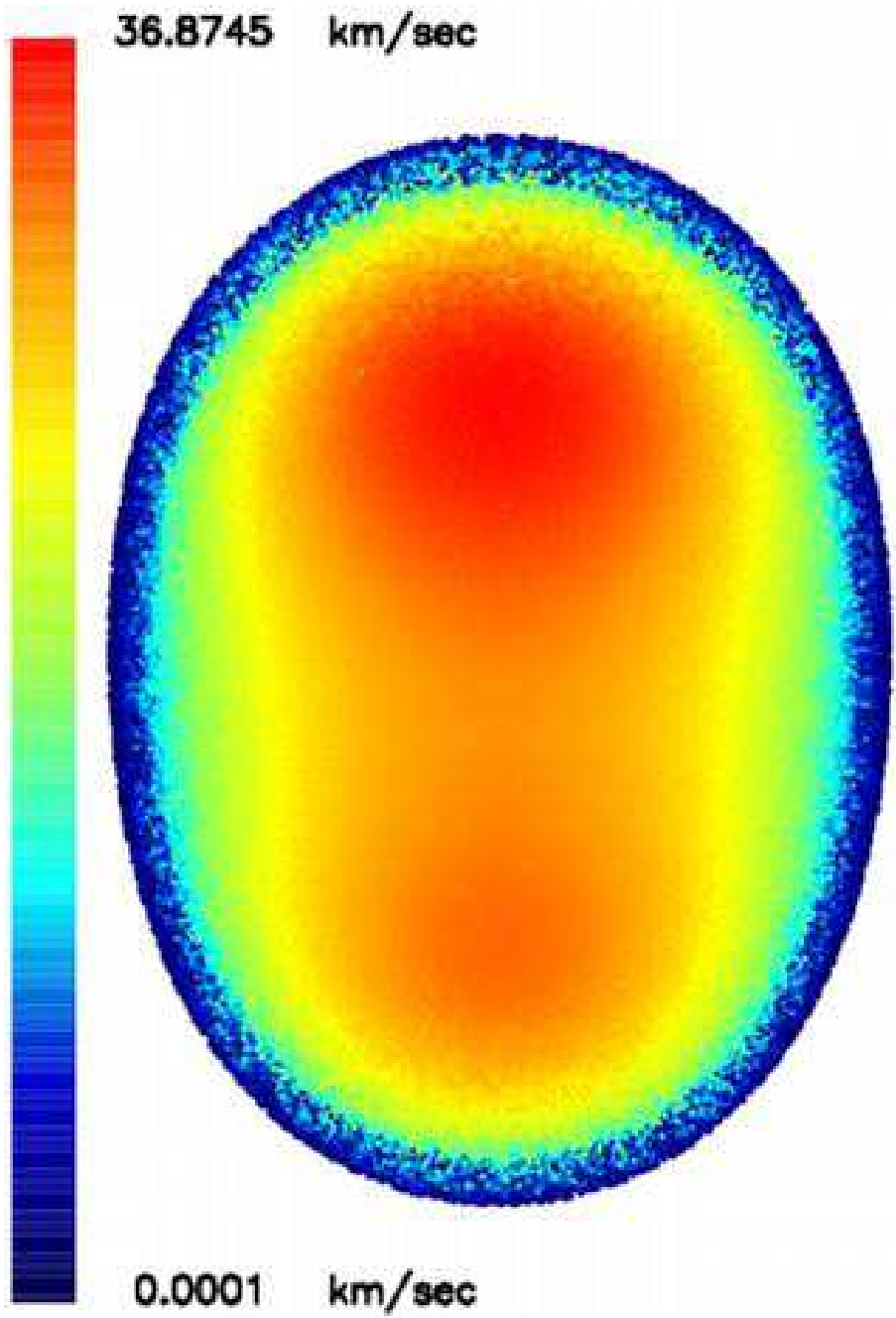}
\caption
{
 Map of the velocity along the line of sight, 
 $V^{\mathrm p}_{\mathrm {theo}}$, 
  relative to a simulation of GW~46.4+5.5~ 
  when 190000 random points are selected on the surface.
  The   physical parameters are the  same as in 
  figure~\ref{worm_velocita_y}.
}
\label{worm_velocita_y}%
    \end{figure}

Perhaps it  is useful  to map the velocity ($V^{\mathrm p}_{\mathrm {theo}}$)
 in the  {\it y}-direction,  
\begin {equation}
V^{\mathrm p}_{\mathrm {theo}} = v (\theta,\phi) \cdot  \sin (\theta) \cdot (\sin \, \phi)
 .
\end   {equation}
This is the velocity measured along the line  of sight 
when an observer stands in front of the super-bubble;
$\theta$  and $\phi$  are defined in subsection~\ref{sectheta}.
The  structure of the projected velocity, $V^{\mathrm p}_{\mathrm{theo}}$, 
is  mapped
in figure~\ref{worm_velocita_y}~
by using different colours; 
the range  is 0~$\mbox{km~s}^{-1}$ 
$\leq~V^{\mathrm p}_{\mathrm {theo}}~\leq$ 36~$\mbox{km~s}^{-1}$
(0~$\mbox{km~s}^{-1}$ 
$\leq~V^{\mathrm {p}}_{\mathrm {theo}}~\leq$ 32~$\mbox {km~s}^{-1}$
with the Euler process)
and  the averaged  projected velocity  is  
$\approx 20 \, \mbox {km~s}^{-1}$ 
($\approx 17.6 \, \mbox {km~s}^{-1}$ with the Euler process),
a value  that is greater by  $\approx 5\,\mbox {km~s}^{-1}$ 
($\approx 2.6 \mbox {km~s}^{-1}$ with the Euler process) than the already
mentioned  observed expansion velocity,
$V_{\mathrm {exp}}=15\,\mbox{km~s}^{-1}$.  

As is evident from the map in   figure~\ref{worm_velocita_y},
the projected velocity of expansion is not uniform
over all of the shell's surface,
  but it is greater in the central
region in comparison  to the external region.
In this particular case of an  egg-shape,
we observe 
a  nearly circular region connected 
with the maximum velocities in the upper part of the shell.
It is  therefore possible to speak of egg-shape 
appearances 
in the 
Cartesian  physical coordinates  
and  spherical-appearances in the projected maximum velocity.

\begin{figure}
\FigureFile(120mm,120mm){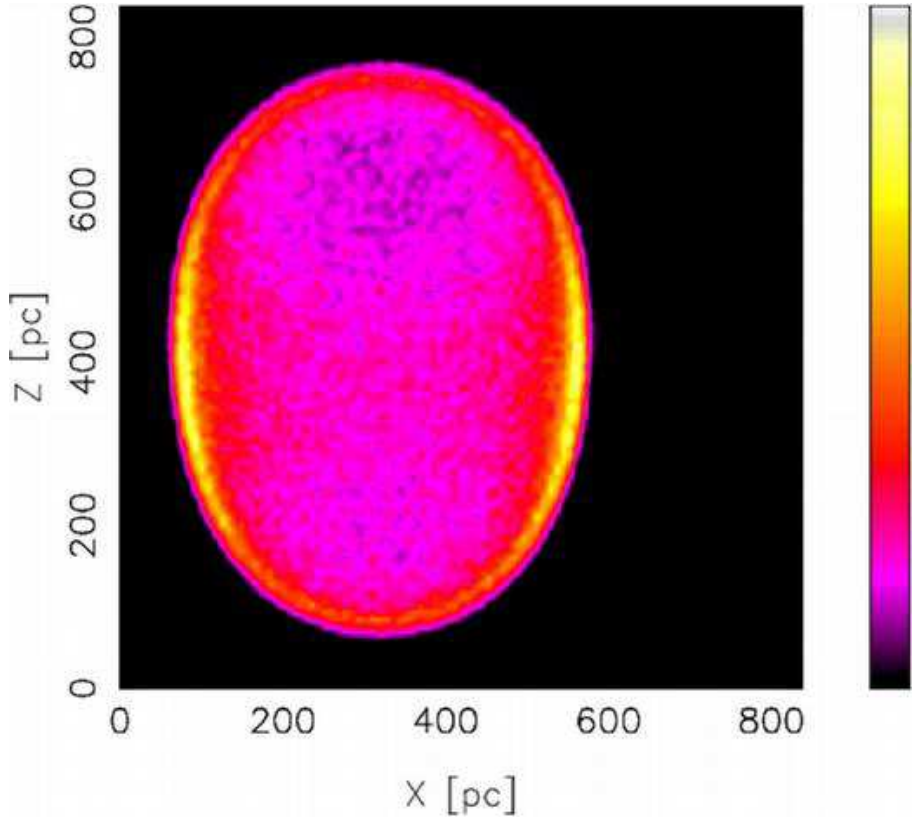}
\caption
{
 Number of particles, $N^{\mathrm {p}}$,   
along the line of sight
 relative to the simulation of GW~46.4+5.5~ when NDIM=101.
 The other parameters are as  in figure~\ref{worm_velocita_y}.
}
\label{worm_ring}%
    \end{figure}

Another way of  presenting  the results is through 
a  ring-enhancement    due
to the law of  e.m. emission in the external layer (see  
appendix~\ref{sec_ring}).  
The projected number of particles  $N^{\mathrm {p}}(i,j)$,
is  then mapped
using  colour-coding contours (figure~\ref{worm_ring});
the  density enhancement  in the ring region~
is  evident from a visual inspection of this simulated map.

\begin{figure}
\FigureFile(120mm,120mm){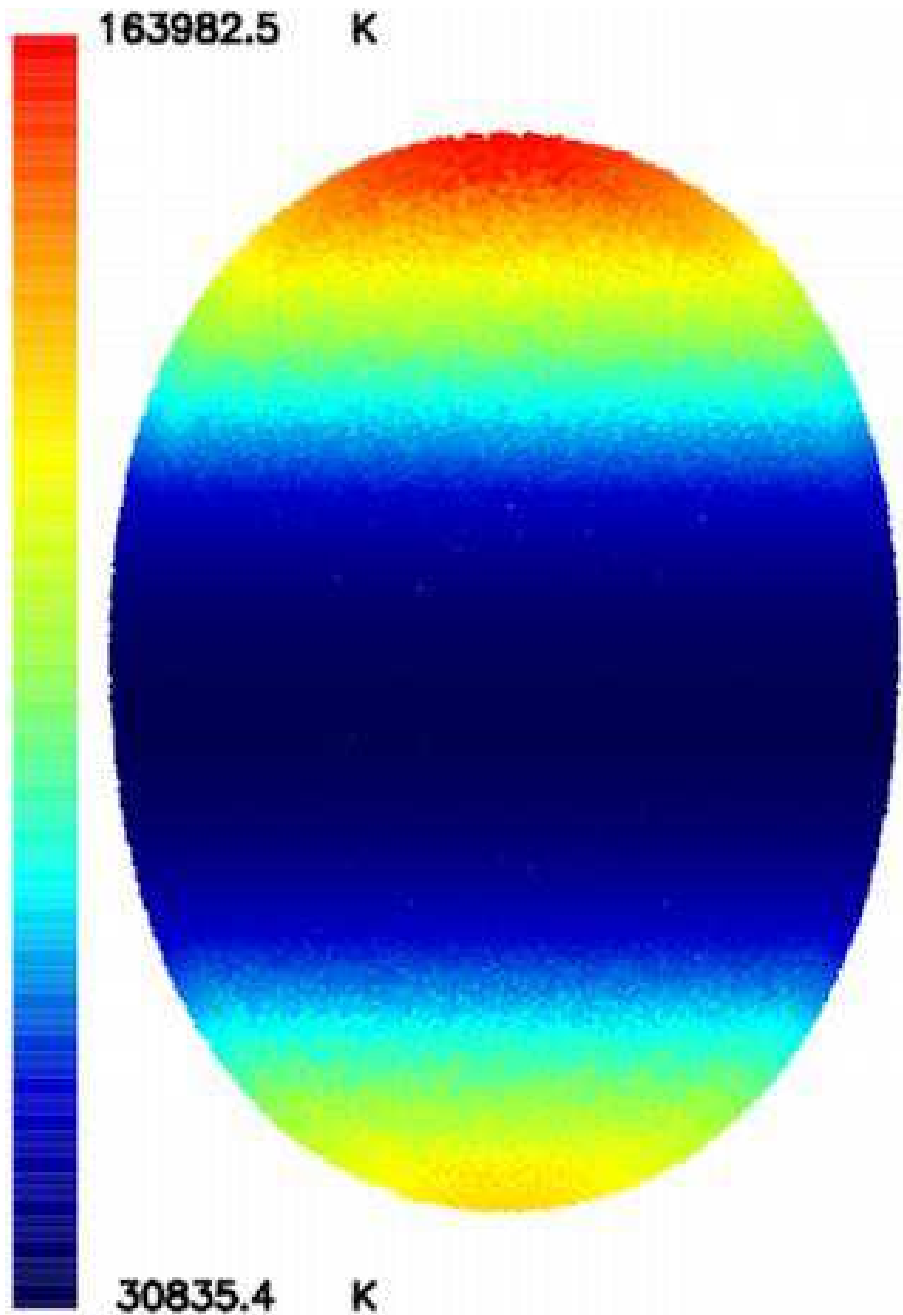}
\caption
{
  Map of the temperature  relative to the simulation 
  of the supershell associated with GW~46.4+5.5;
  the   other parameters are the  
  same as in figure~\ref{worm_velocita}.
}
\label{temperature_worm}%
    \end{figure}

Upon  adopting formula~(\ref{formulat}), it is  
 possible to map the field
of temperature  of the supershell associated with GW~46.4+5.5
(see figure~\ref{temperature_worm}).

\begin{figure}
\FigureFile(120mm,120mm){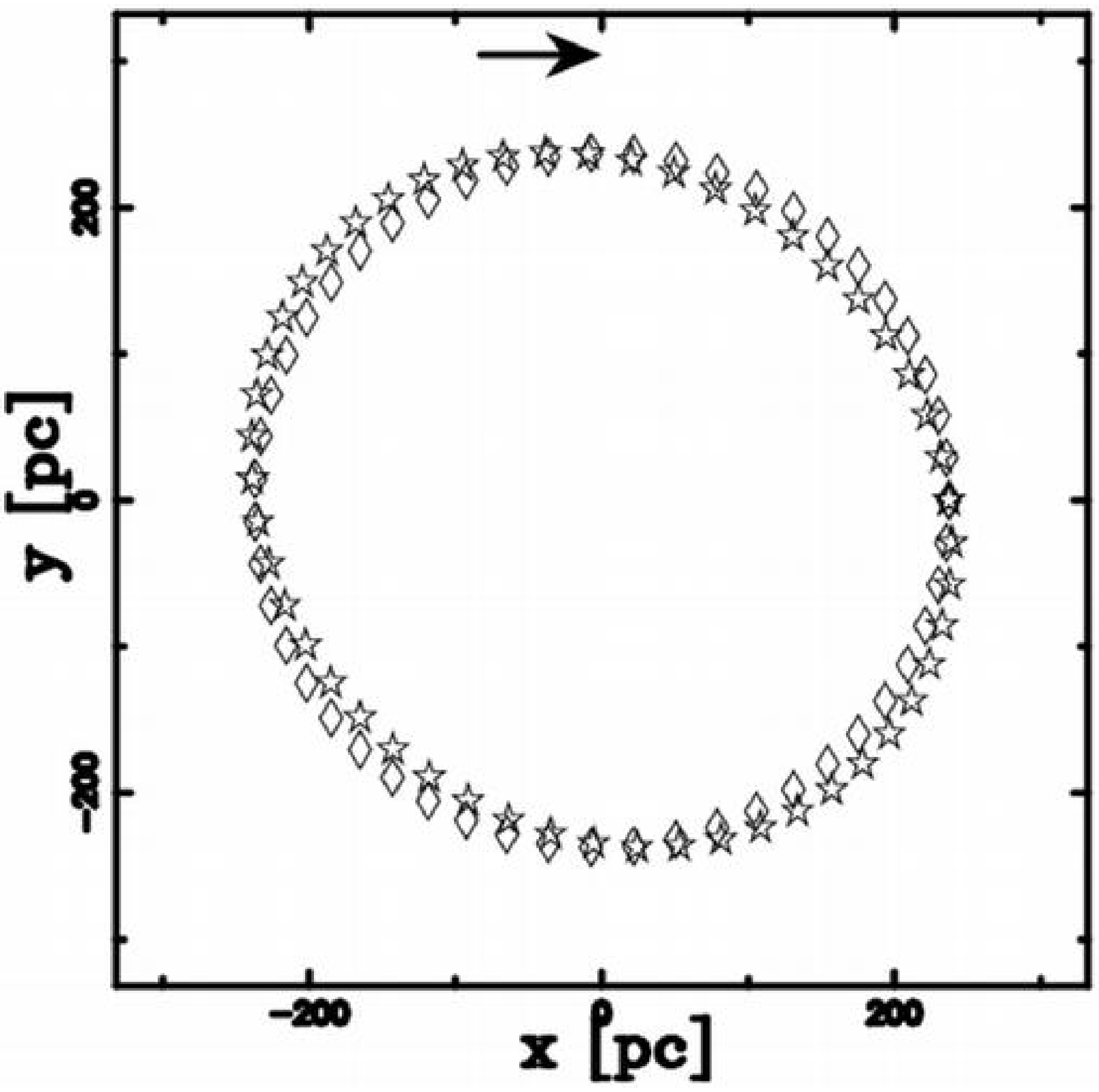}
\caption
{
Circular and rotation-distorted sections of 
the superbubble on the {\it  x-y}   plane, 
denoted by  {\it z}=$z_{\mathrm{OB}}$. The
physical parameters are the  same as in figure~\ref{worm};
the rhombi represent the circular  section and the stars 
the rotation-distorted section; the Galaxy direction of 
rotation is also  shown.  
}
\label{rotation_worm}%
    \end{figure}

The effect  of  galactic rotation on the shape  of the superbubble
is shown in figure~\ref{rotation_worm}
where a circle is transformed into  an ellipse.

\subsection{GSH 238 }

\label{sec238}
  The physical 
parameters concerning  GSH\,238,   as  deduced in 
\citet{heiles},
  are reported in  table~\ref{tab:gsh238}.
   \begin{table}
      \caption{Data of the super-shell associated with GSH\,238.}
         \label{tab:gsh238}
      \[
         \begin{array}{cc}
            \hline
            \noalign{\smallskip}
\mbox {Size (pc}^2)                    & 440 \cdot 1100 \mbox{~at~b=0} \\
\mbox {Expansion~velocity~(km~s}^{-1})  & 8                \\
\mbox {Age~(10}^7\,{\mathrm{yr}})                  & 2.1              \\
\mbox {Total energy\,(10}^{51}{\mathrm{erg}})      & 34               \\
            \noalign{\smallskip}
            \hline
         \end{array}
      \]
   \end{table}
The total energy   is obtained  based on 
a model  derived  by  \cite{heiles1979} 
and  the  kinematic  age  is derived from approximate arguments.
Once again, in order to obtain 
$E_{\mathrm{tot}}=34.~10^{51}{\mathrm{erg}}$ 
$[$ see formula~(\ref{etotburst}) $]$ 
and  $t^{\mathrm{burst}}=0.015~10^7~{\mathrm{yr}}$,
we have inserted 
$N^*$=~11300.

\begin{figure}
\FigureFile(120mm,120mm){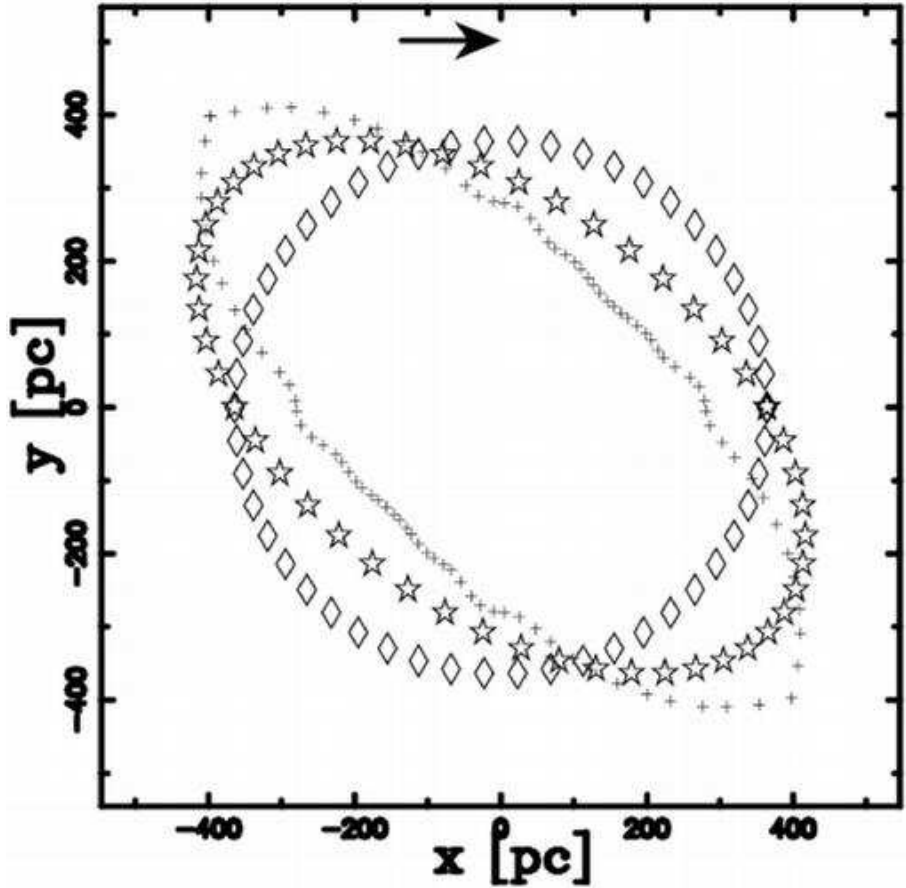}
\caption
{
Rhombi represent the circular  section;
 the stars, 
the rotation-distorted section; and the small crosses,
the observed data of GSH\,238 at {\it b} =0 (our {\it z}=0) 
as extracted from figure 8  
of \citet{heiles}.
The Galaxy direction of  rotation is also  shown.  
The parameters are
$t_\mathrm{\mathrm {age}}$=$2.1\cdot10^{7}~\mathrm{yr}$,  
$\Delta$~$t=0.001\cdot10^{7}~\mathrm{yr}$,
$t^{\mathrm{burst}}_7$= 0.015, $N^*$=  11300, $z_{\mathrm{OB}}$=0 pc,
and $E_{51}$=1.
}
\label{gsh238}%
\end{figure}

The efficiency over 50+1 
directions computed with 
formula~(\ref{eq:reliability}) of a section {\it x-y} when {\it z}=0,
is $\epsilon_{\mathrm {obs}}=58.4\%$.
The 2D  cut at {\it z}=0 of the superbubble  can be
visualised in~figure~\ref{gsh238}.
Our model gives a radial velocity at {\it z} =0  
$V_{\mathrm {theo}}$=6\,$\mbox {km~s}^{-1}$ 
against the observed  8\,${\mbox {km~s}}^{-1}$ at {\it b}=0.  

\section   {Collective Effects}

It should be remembered  that the  worms  have
an average  linear extent  in the {\it z}
 direction  of 150~\mbox{pc}
and  a median width of 70~\mbox{pc};
the maximum extension that  can be reached in the {\it z}-direction is
$\approx$ 1~kpc~(\cite{koo}).
These  observations can be simulated
by  introducing
a coupling  between  the  percolation network  reviewed
in  appendix~\ref{spiral}\/
and the  explosion
 {\em model} described in
  section~\ref{approximation}.
\label{percolation}
The percolation offers:
\begin{enumerate}
\item  The  Cartesian coordinates  of the explosions in
       the galactic plane that follow the grand-designed
        spiral  galaxies.
\item  The times   of the clusters   vary  between
        10$^7$ \mbox{yr} and  10$^8$ \mbox{yr}. This
        cluster age divided by 4 
        can be identified with  $t_{\mathrm{\mathrm {age}}}$,
        the age of the superbubble.
        With this last choice,  the maximum 
        $t_\mathrm{\mathrm {age}}$  is $2.5\cdot10^{7}~\mathrm{yr}$,
        according to the age of GSH\,238, the oldest super-shell
        analysed here.
         In other words, the time sequence of the clusters 
         (1,2,.....,10$\cdot 10^{7}~\mathrm{yr}$) is converted 
         in the following age sequence  of the 
         super-shells (0.25,0.5,.....,2.5~$\cdot 10^{7}~\mathrm{yr}$).
         This  choice fixes the maximum altitude  reached by the 
         super-shells in $\approx$ 600 ~pc.

\end{enumerate}
In order to test the suggested coupling between  superbubbles
and the percolation 
the  distance of 200~pc,  which    the explosions should cover
in one time step, $10^7~{\mathrm{yr}}$ was  inserted  in 
equation~(\ref{eqn:raadia}), in order to trigger
new cluster formations.

\begin{figure}
\FigureFile(120mm,120mm){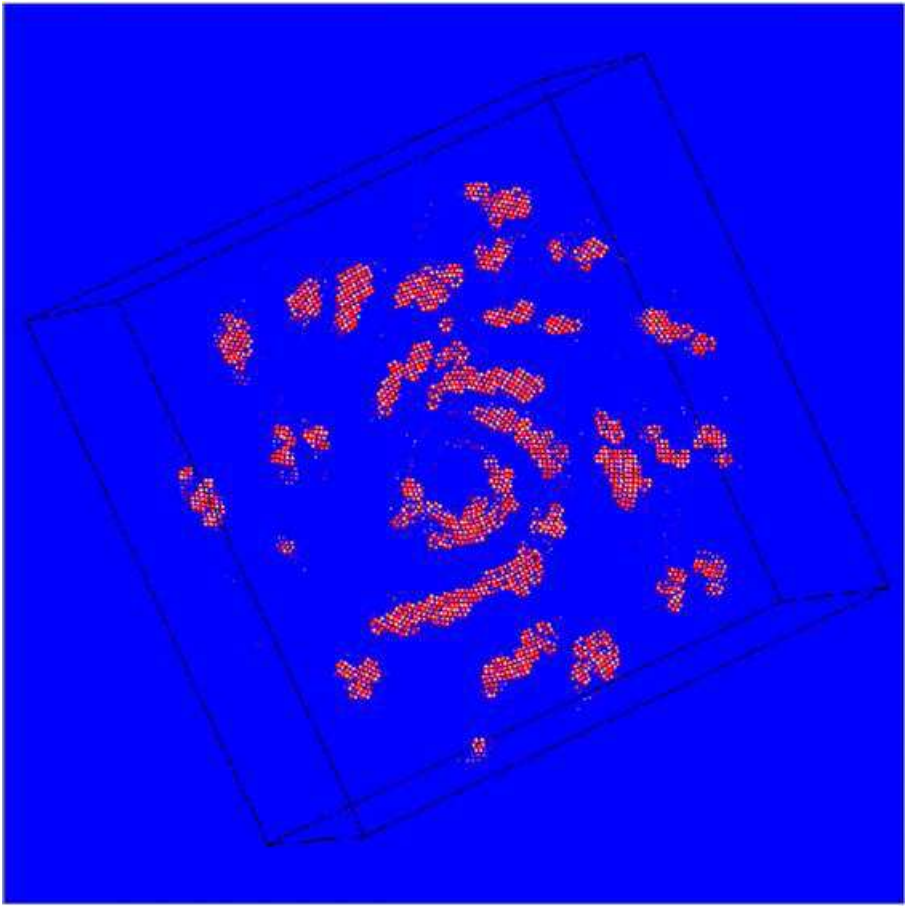}
\caption
{
Simulation of spiral galaxy $S_{\mathrm {b}}$  by adopting the
theory and the parameters of appendix~\ref{spiral}.
The clusters  are represented by   spheres of
decreasing radius with increasing age.
The three Eulerian angles
characterising the point of view of the observer
are  $ \Phi $= 25$^{\circ }$
 , $ \Theta $= 75$^{\circ }$, 
 and  $ \Psi $= 25$^{\circ }$.
}
\label{figperc}%
    \end{figure}

Upon  adopting the standard values of $n_0=0.4 \mbox {particles~cm}^{-3}$,
$E_0$=1 and  $N^*$=150, we obtain $t^{\mathrm{burst}}_7$=0.035,
 a value
that seems to be in agreement with our previous assumption in
which  the bursting \underline{phase} is lower than the age
of the superbubble.
A typical percolation run is shown in figure~\ref{figperc}
for a face-on galaxy.
The galactic rotation is considered in all the simulations
discussed  in this section.

\subsection{Few Superbubbles}

\label{sec_few}
\begin{figure}
\FigureFile(120mm,120mm){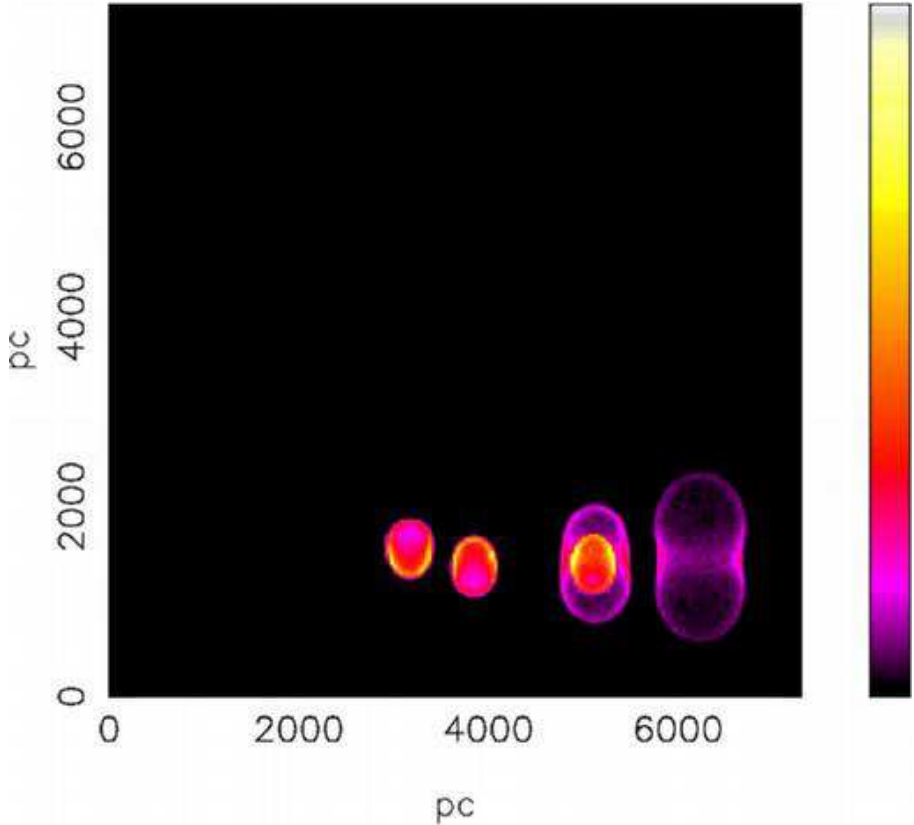}
\caption
{
 Number of particles, $N^{\mathrm {p}}$,   along the line of sight
 relative to the   network   of the explosions when
 NDIM=200 and the galaxy is situated edge on.
 The  boundaries of the  box are  XMIN=YMIN=-8000 \mbox{pc},
 XMAX=YMAX=-4000 \mbox{pc};
  5 clusters were   selected and each superbubble
  had 40000  random points on it's  surface.
The parameters  were 
$\Delta t=0.01 \cdot 10^{7}~\mbox{yr}$,
$t^{\mbox{burst}}_7$= 0.2,  and  $N^*$=  50.
}
\label{plane_sb}%
    \end{figure}

In order  to visualise a few   explosions of coordinates
{\it X}  and {\it  Y},
only those  clusters comprised  between 
XMIN[\mbox{pc}] $\Leftrightarrow$ XMAX[\mbox{pc}]
and  YMIN[\mbox{pc}] $\Leftrightarrow$ YMAX[\mbox{pc}]
should be selected.
The distance  $z_{\mathrm{OB}}$ from the galactic plane
of the clusters  is  allowed to randomly vary
 between $-$100 pc
and 100 pc, which  is the size of the active regions.
A  typical example  of this network  of
multi-explosions  is  shown  in figure~\ref{plane_sb}
through the algorithm that counts  the number of particles
along the line of sight; the selected
boundaries    are marked  on  plots and the
appearance of the rings  is evident.
%

\begin{figure}
\FigureFile(120mm,120mm){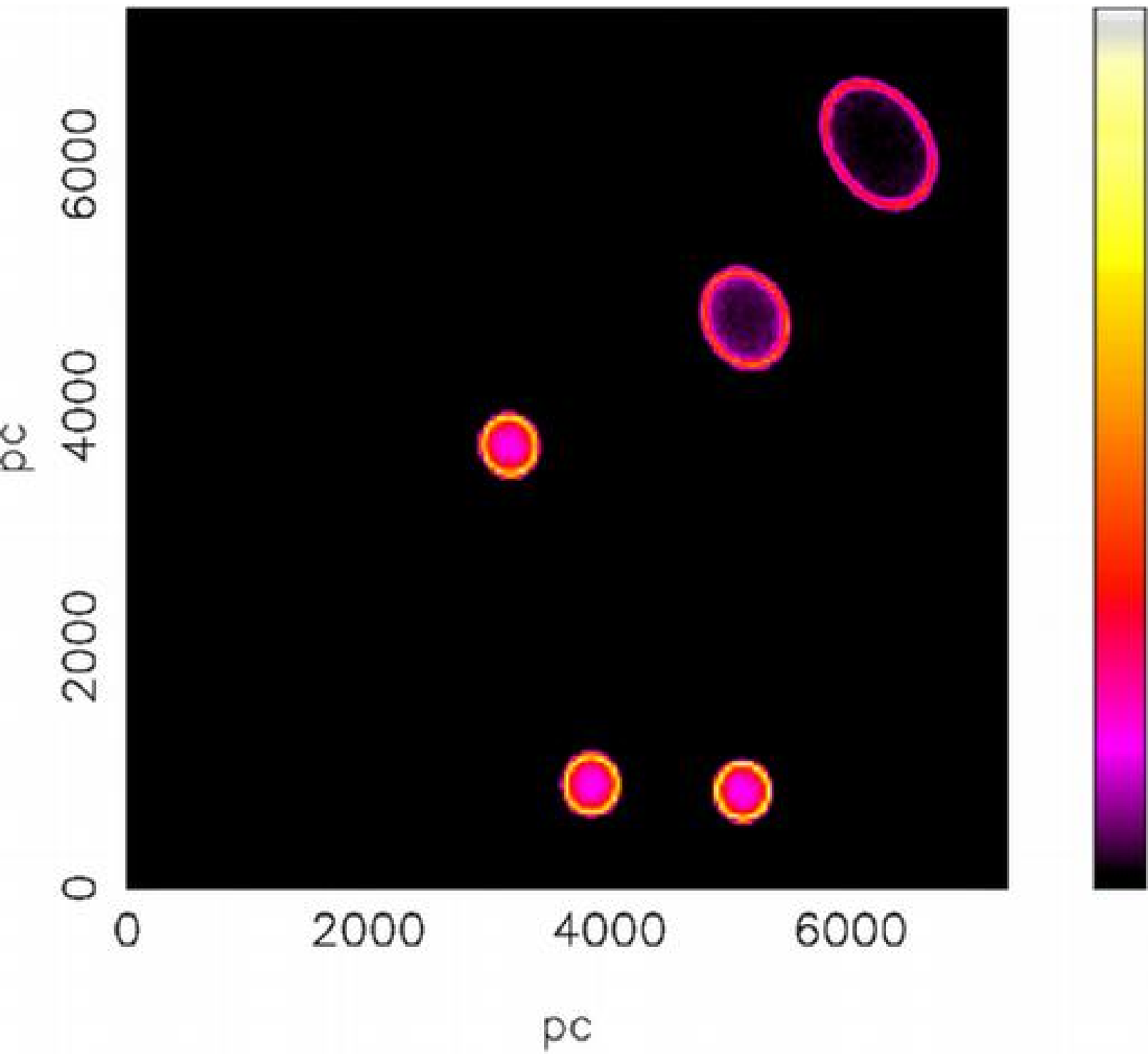}
\caption
{
 Same as figure\,\ref{plane_sb},  but 
 when the galaxy is situated face  on.
}
\label{plane_sb_front}%
    \end{figure}

The superbubbles  can also be visualised when the observer is situated
in front of the galaxy plane (see  figure~\ref{plane_sb_front}).
%
The effect of shape distortion due  to formula~(\ref{phitotale})
is evident.

\subsection{The Galactic Plane}

\label{sec_many}
\begin{figure}
\FigureFile(120mm,120mm){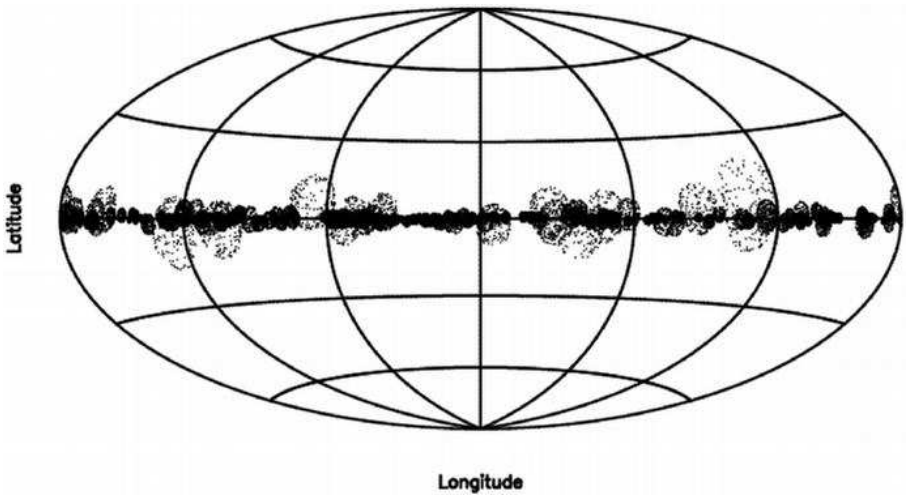}
\caption
{Structure of  the galactic plane in the Hammer-Aitof  projection,
as resulting from the superbubble/percolation  network.
The value  of  pselect  is 0.08,  corresponding to 260  selected
clusters.
The parameters   were 
$\Delta t=0.01 \cdot 10^{7}~\mbox{yr}$,
$t^{\mbox{burst}}_7$=0.5, $N^*$=  100 and each superbubble had
200 random points on its surface.
}
\label{plane_hamm}%
    \end{figure}

In order  to simulate the structure
of the galactic plane,
 the same  basic  parameters  as in
figure~\ref{figperc} can be chosen,
but now    the superbubbles are drawn   on
an equal-area Aitof projection.
 In particular,
a certain  number of  clusters
 will be selected  through  a random process
  according  to the following formula:
\begin{equation}
selected~clusters = pselect  \cdot  number~of~clusters~from ~percolation
, 
\end{equation}
where   pselect  has  a  probability  lower than one.
The  final  result  of the  simulation
is reported   in figure~\ref{plane_hamm}.

\begin{figure}
\FigureFile(120mm,120mm){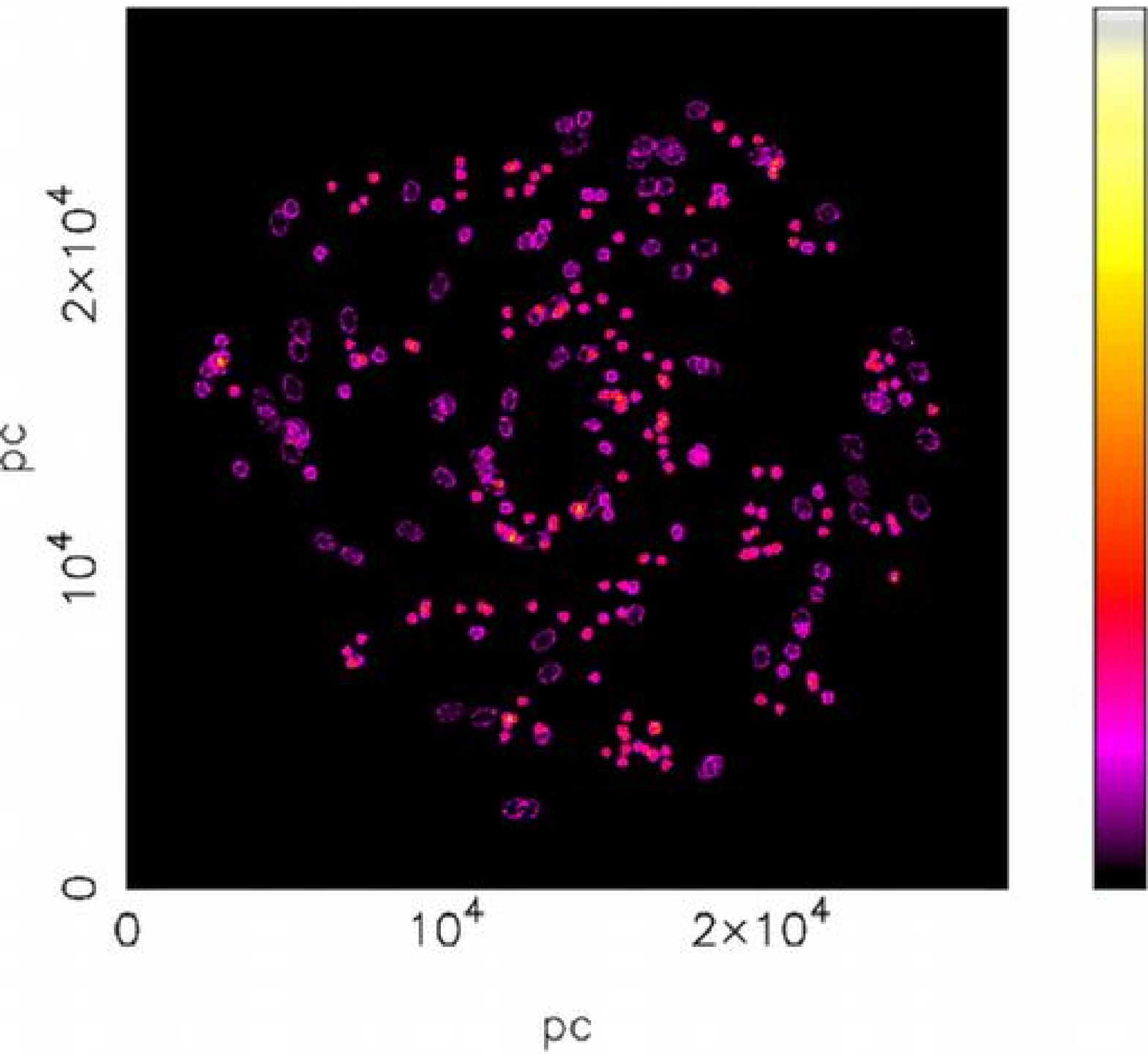}
\caption
{
 Number of particles, $N^{\mathrm {p}}$,   along the line of sight
 relative to the   network   of the explosions  when
 the galaxy is face on. Physical  parameters  were as in
 figure~\ref{plane_hamm} and NDIM= 401.
}
\label{spiral_bubbles}%
    \end{figure}

We can also visualise the structure of the superbubbles
as seen from an observer situated outside the Galaxy;
the spiral structure arising from the
ensemble of the shells is evident  (see figure~\ref{spiral_bubbles}).
The elliptical  shape of the superbubble, 
according to 
formula~(\ref{phitotale}), is a function of both the age  and the distance
from the center. The distortion also follows the same
inclination of the arms.

\section{Conclusions}

The expansion of a super-bubble in the ISM belonging to our galaxy 
can be simulated by applying   Newton's second law 
to  different pyramidal  sectors. 
The following results  were  achieved:

\begin{enumerate}
\item      Characteristic structures,  such as    hourglass-shapes, 
           vertical walls and  V-shapes,  
           could  be obtained by varying the basic parameters, 
           which  are   the bursting \underline{phase} and the time 
           over which the phenomena  are followed. 
\item      Single objects,  like the super-shells associated 
           with GW~46.4+5 and GSH\,238  were simulated with an
           efficiency $\epsilon_{\mathrm {obs}}\approx 68 \%$
            and $\epsilon_{\mathrm {obs}}\approx 58  \%$, respectively.
\item      The network of many explosions that originate from 
           the galactic plane could  be tentatively 
           simulated. 

\end{enumerate}
Some problems are  still  unsolved:  
\begin {enumerate}
\item The results  were obtained by using a standard law concerning 
      the dependence of the ISM  on  the distance from
      the galactic plane, {\it z}. Perhaps  a more refined
      treatment of the H\,I thin disk will change  the obtained 
      morphologies.
\item  Recent maps based on  measurements  of Na\,I absorption
       in the surroundings of the  Local
       Bubble suggest the existence of narrow 
       tunnels (\cite{lallement}).
       Can   a tunnel be explained 
       by the  interaction of a super-bubble arising 
       from the galactic plane (GSH~238)
       with a smaller interstellar bubble (the Local Bubble)~?
\end {enumerate}

\appendix
\section{The Density Projection Algorithm }

\label{sec_ring}
Consider a  shell within two concentric
spheres with  radii {\it R} and   $R+\Delta R$.
The total intensity of  thermal e.m.  radiation, 
when absorption and
scattering  are neglected,
is proportional to the following parameter (chosen among
others): $I \propto  ~l~$,
where  {\it l }
is the dimension of the radiating region along the line
of sight; $I_{\mathrm{ring}}$  and  $I_{\mathrm {center}}$ denote,
respectively,  the intensity along the
ring and  along the center.
The  length of  the emitting
region when  the line of sight  touches  the inner shell
is
\begin{equation}
l_{\mathrm{ring}}=2\sqrt{(R+\Delta R)^2-R^2}
.
\end{equation}
 Conversely,
when  the line of sight  is aligned  with the center of
the two  shells the  emitting length  is
\begin {equation}
l_{\mathrm{center}}=2 \Delta R
  .
\end{equation}
We briefly remember that  on applying  mass conservation
to a region of radius {\it R} before
and after an  explosion and the junction condition
concerning the density,
the following formula concerning
the depth of the expanding layer  is obtained:
\begin {equation}
\frac {\Delta R}{R} =
\frac {\gamma -1 }{\gamma+1} \frac {1}{3} = \frac {1}{12}
,
\end   {equation}
where $\gamma=\frac{5}{3}$  (see ~\cite{deeming}).
By using a little algebra,  we find
\begin {equation}
\frac {l_{\mathrm{ring}}} {l_{\mathrm{center}}} = 5
  .
\end {equation}
This  argument  is  easily  generalised in  the case
of  $\Delta R=R/n_{\mathrm {t}}$   where  $n_{\mathrm {t}}$ is  an integer
greater than  12 :
\begin {equation}
\frac {l_{\mathrm{ring}}} {l_{\mathrm{center}}} \approx  \sqrt {2 n_{\mathrm {t}}}
 \mbox { when  }
 \Delta~R=R/n_{\mathrm {t}}
  .
\end {equation}
\label {darkening}
Therefore, the ratio $\frac{ I_{\mathrm{ring}}} {I_{\mathrm{center}}} =
\frac{ l_{\mathrm{ring}}} {l_{\mathrm{center}}}$ can be  called "the center darkening
law", and says  that the center of the superbubble
should have a lower emergent intensity than the shell  region
(the limb).
We point out     that  on the expanding surface
of the superbubble there is
a certain  value  of  the intensity  $[$for example,  30
$\frac {\mathrm MJy}{\mathrm sr}$ in the band of 100\,$\mu m$~(far-IR) (see
figure~1~(gray-scale intensity) in~\citep{kim})$]$.
The  intensity at the center of the superbubble
is theoretically  predicted to  be  approximately five times
lower  than that of the ring  
(therefore, 5 $\frac {\mathrm MJy}{\mathrm sr}$).
This argument was applied to our superbubble
by plotting the  density  of random particles
projected  on the sky.  In order to implement  such a projection,
we first computed  the number of random elements
($N(i,j,k)$) that   fall  in a little cube of  volume
$ \mathrm {side^3  / NDIM^3 }$ where   {\it side}  is considered
to be  the length of
the cube that encloses the superbubble,  and NDIM the number
of considered pixels.
We then sum over one index, for example k, in order to
obtain  the elements  $N^{\mathrm p}(i,j)$   that  fall in a given
$\mathrm {area=\frac{side^2}{NDIM^2}}$. 
The already described algorithm
can be considered a simulation of the number of particles
that are being emitted  on the line of sight.

\section{Percolation and spiral galaxies}

\label{spiral}
The appearance of  arms can be simulated through
the percolation
theory \\
~\citep{seiden1,seiden2,seiden3,seiden4};
  here, a  previous  model~\citep{zaninetti88}
will be  slightly modified.
The fundamental   hypotheses
and   the parameters adopted in the
simulation are now reviewed:
\begin{enumerate}
\item
The motion of  gas on the galactic plane has a constant
rotational  velocity, named
${\it V}_{\mathrm{G}}$ (in the case of
spiral type Sb     218 $\mathrm {km\,s^{-1}}$).
Here the velocity, ${\it V}_{\mathrm {G}}$, is
expressed in 200 $\mathrm {km\,s^{-1}}$~units and will therefore be
${\it V_{\mathrm{G}}}$ =1.09.
\item
The polar simulation array   made by rings and cells
has  a radius
${\it R}_{\mathrm{G}}$ = 12~\mbox{kpc}.
The  number of rings, (60),
  is  the multiplication together   of  ${\it R}_{\mathrm{G}}$
and  the number of rings for each
kpc, named  \mbox{nring/kpc},
which in our case is  five .
Every ring is then made up of
many~{\it cells}, each one with a  size on  the
order of the galactic thickness,
 $ \approx$ 0.2 \mbox{kpc}.
The parameter  \mbox{nring/kpc} can  also be found
by
dividing {\mbox  1kpc} by the approximate cell's size.

\item  The global number of cells, 11121,
multiplied by the probability
of spontaneous new cluster formation,
$p_{\mathrm {sp}}$ (for example 0.01),
 allows the process to start (with the previous parameters,
 111 new clusters were generated).
Each one of these
sources has six new surroundings that are labelled for each ring.
\item   In order to better simulate the decrease of the gas density
along the radius,  a stimulated  probability of forming
new clusters
with  a linear dependence by  the radius
,$ p_{\mathrm st}$ = $ a + b R $, was chosen.
The values a and b are found by fixing ${\it prmax}$
( for example 0.18),
 the stimulated probability
at the outer ring, and ${\it prmin}$  (for example 0.24)
 in the inner ring;
of course,
{\it prmin}$\geq ${\it prmax}.
This  approach  is surprisingly similar to the introduction
of  an anisotropic
probability distribution in order to better
simulate certain classes of spirals~\citep{jungwiert}.
\item   Now, new sources are selected in
each ring based on  the hypothesis of  different
stimulated probabilities.
A rotation curve is imposed
so that the array rotates in the same  manner
as the  galaxy . 
The
procedure repeats
itself n times (100); we denote this  by ${\it t}_{\mathrm{G}}$ the
age  of the simulation,$100\cdot10^7 \mbox {yr}$
being  $10^7$ \mbox{yr},  the
astrophysical counterpart of one time  step.
\item In order to prevent catastrophic growth, the process
is stopped when the number of surroundings is
greater than ${\it max}$ (1000)
and restarts by spontaneous probability.
\item The final number of active cells  (3258)
is plotted with the size, 
which  decreases linearly  with the cluster age.   In
other words,  
the young cluster are bigger than the old ones.
Only ten
cluster ages  are shown;
only  cells with an
age of less than ${\mathrm{ life}}$  (in our case $10\cdot10^7$ yr)
are selected.
A more deterministic  approach has been followed 
by~\citet{palous94},
where  a model that connects 
the energy injection by star  formation with the resulting
interstellar  structures in a differentially rotating disc  has been
introduced. 

\end{enumerate}


\begin{thebibliography}{}

\bibitem  [Basu et al. (1999)]{basu} Basu, S.,
      Johnstone, D., \& Martin, P.~G.  1999, \apj, 516, 843

\bibitem[Begelman, Li(1992)]{begelman} Begelman, M.~C.~,\& Li, 
                    Z.-Y.\ 1992, \apj, 397, 187 

\bibitem[Bisnovatyi-Kogan, Silich(1995)]{bisnovatyi} 
              Bisnovatyi-Kogan, G.~S.~,\& Silich, S.~A.
              \ 1995, Rev.  Mod. Phys., 67, 661 

\bibitem[Boulares, Cox(1990)]{boulares} Boulares, A.~,\& Cox, 
                D.~P.\ 1990, \apj, 365, 544 

\bibitem [Deeming, Bowers (1984)] {deeming}
      Deeming, T.,\& Bowers, R.L. 1984,
      Astrophysics II: Interstellar Matter and Galaxies,
      (Boston: Jones \& Bartlett Pub)


\bibitem [Dickey,  Lockman (1990)]{dickey}
               Dickey, J.M. ,\&  Lockman, F.J.  1990,
               ARA$\&$A, 28, 215

\bibitem[Ferriere et al. (1991)]{ferriere}
Ferriere, K.~M., Mac Low, M., \& Zweibel, E.~G.\ 1991, \apj, 375, 239.

\bibitem[Franco et al. (2002)]{franco2002} 
       Franco, J., Kim, J., Alfaro, E.~J., \& Hong, S.~S.
       \ 2002, \apj, 570, 647 

\bibitem [Hartmann, Burton (1997)]{hartmann}
            Hartmann,D. ,\& Burton, W.B.  1997,
            Atlas of Galactic Neutral Hydrogen (Cambridge:
            Cambridge University Press)
\bibitem[Heiles(1979)]{heiles1979} Heiles, C.\ 1979, \apj, 229, 
         533 
\bibitem[Heiles(1998)]{heiles} Heiles, C.\ 1998, \apj, 498,
            689
\bibitem [Igumenshchev et al. (1990)]{igu}
            Igumenshchev, L.V.,Shustov, B.M. ,\& Tutukov, A.V. 1990,
            A\&A , 234, 396
\bibitem  [Jungwiert , Palous (1994)]{jungwiert}
           Jungwiert, B. ,\& Palous,J. 1994,  A\&A,   287,55
\bibitem  [Kamaya (1998)]{kamaya} 
          Kamaya, H.  1998, \apjl, 493, L95

\bibitem[Kim et al.(2000)]{franco}k Kim, J., Franco, J., Hong, 
          S.~S., Santill{\' a}n, A., \& Martos, M.~A.
          \ 2000, \apj, 531, 873 

\bibitem  [Kim , Koo (2000)]{kim}
          Kim,K.T. ,\& Koo,B. C. 2000, ApJ , 529, 229

\bibitem  [Koo, McKee (1990)]{koo0} Koo, B. C.~,\& McKee, C.~F.\
           1990, \apj, 354, 513

\bibitem  [Koo et al. (1992)]{koo}
          Koo,B.C.,Heiles,C. ,\& Reach,W.T 1992, ApJ , 390 , 108
\bibitem [Lockman (1984)]{lockman}
          Lockman,  F.J.  1984, ApJ , 283, 90

\bibitem[Mac Low, McCray (1988 )] {mac0} Mac Low, M.-M. ,\&
         McCray, R.\ 1988, \apj, 324, 776.

\bibitem [MacLow et al. (1989)]{mac}
         MacLow, M.-M. , McCray, R.  ,\& Norman,M.L. 1989, 
         ApJ , 337,141
\bibitem  [McCray (1987)]{mccray}
       McCray,  R.  1987, 
       in Spectroscopy of Astrophysical Plasmas,
       ed.\  A. Dalgarno A. \& D. Layzer  (Cambridge:
       Cambridge University Press)  271
\bibitem  [McCray , Kafatos (1987)]{mccrayapj87}
    McCray, R. ,\& Kafatos , M.  1987, ApJ , 317,190

\bibitem [ McKee (1987)]{mckee}    
          McKee C.F., 1987, 
       in Spectroscopy of Astrophysical Plasmas,
       ed.\  A. Dalgarno A. \& D. Layzer  (Cambridge:
       Cambridge University Press)  226

\bibitem[Moreno et al. (1999)]{moreno} Moreno, E., 
    Alfaro, E.~J., \& Franco, J.\ 1999, \apj, 522, 276 

\bibitem[Palous et al. (1990)]{palous1990} 
       Palous, J., Franco, J., ,\& Tenorio-Tagle, G.\ 1990,
        \aap, 227, 175 

\bibitem[Palous et al (1994)]{palous94} 
        Palous, J., Tenorio-Tagle, G., \& Franco, J.
        \ 1994, \mnras, 270, 75


\bibitem[Pikel'ner(1968)]{pikelner} 
        Pikel'ner, S.~B.\ 1968, 
        \aplett, 2, 97 

\bibitem [Press et al. (1992)]{press}
   Press,W. H.          ,
   Teukolsky, S. A.     ,  Vetterling, W. T. , 
   \&  Flannery,   B. P.,  1992
   Numerical Recipes in  Fortran , second edition
       (Cambridge:
       Cambridge University Press) ,
        117

\bibitem  [Puche et al. (1992)]{puche}
       Puche,D.,Westpfahl,D.,Brinks,E. ,\& Roy,J.-R. 1992,
       AJ  , 103, 1841
\bibitem[Santill{\' a}n et al. (1999)]{santillan} 
       Santill{\' a}n, A., Franco, J., Martos, M., \& Kim, J.
       \ 1999, \apj, 515, 657 

\bibitem[Schulman, Seiden (1986)]{seiden3} Schulman, L.~S.~,\&
         Seiden, P.~E.\ 1986, Science, 233, 425

\bibitem[Seiden (1983)]{seiden2} Seiden, P.~E.\ 1983, \apj, 266,
        555

\bibitem[Seiden, Gerola (1979 )]{seiden1} Seiden, P.~E.~,\&
Gerola, H.\ 1979, \apj, 233, 56

\bibitem [Seiden , Schulman (1990)]{seiden4}
        Seiden,P.E. ,\& Schulman,L.S.  1990, 
        Advances in Physics , 39,1

\bibitem[Silich (1992)]{silich4} Silich, S.~A.\ 1992, \apss,
        195, 317 



\bibitem [Silich et al. (1996)]{silich} Silich, S.\ A., Franco, J.,
        Palous, J. ,\& Tenorio-Tagle, G.  1996, ApJ, 468, 722

\bibitem [Silich et al. (1996)]{silich96} 
         Silich, S.~A., Mashchenko, S.Y. , 
         Tenorio-Tagle, G., ,\& Franco, J.\ 1996, \mnras, 280, 711 


\bibitem[Stone, Norman(1992)]{stone} Stone, J.M.~,\& 
         Norman, M.~L.\ 1992, \apjs, 80, 753 


\bibitem [Tenorio-Tagle, Bodenheimer(1988)]{tenorio} 
       Tenorio-Tagle, G.~,\& Bodenheimer, P.\ 1988, 
       \araa, 26, 145 


\bibitem[Tomisaka (1992)]{tomisaka2} Tomisaka, K.\ 1992, \pasj ,
        44, 177.

\bibitem[Tomisaka (1998)]{tomisaka3} Tomisaka, K.\ 1998, \mnras ,
        298, 797.

\bibitem [Tomisaka, Ikeuchi (1986)]{tomisaka}
        Tomisaka,K. ,\& Ikeuchi,S. 1986, PASJ, 38,697


\bibitem [Walter et al. (1998)]{walter}
            Walter, F., Kerp, J.,
            Duric, N., Brinks, E. ,\&  Klein, U.
            1998, \apj  , 502, L143
\bibitem[Weaver et al.(1977)]{weaver} 
         Weaver, R., McCray, R., 
         Castor, J., Shapiro, P., 
         ,\& Moore, R.\ 1977 , \apj , 218, 377 

\bibitem[Welsh et al.(2003)]{lallement} Welsh, B.~Y., Lallement,
         R., Vergely, J.~L., Crifo, F., \& Sfeir, D.\ 2003,
         BAAS, 202,53 


\bibitem[Wouterloot et al. (1990)]{wouterloot} 
         Wouterloot, J.~G.~A., Brand, J., Burton, 
         W.~B., \& Kwee, K.~K.\ 1990, \aap, 230, 21 

\bibitem [Zaninetti (1988)]{zaninetti88}
        Zaninetti, L.  1988,   A\&A , 190, 17

\end{thebibliography}
\end{document}